\documentclass[10pt]{article}

\usepackage{amsmath}
\usepackage{amssymb}

\usepackage{graphicx}

\usepackage{cite}

\usepackage{color}

\usepackage{setspace}
\usepackage[labelfont=bf,labelsep=period,justification=raggedright]{caption}

\makeatletter
\renewcommand{\@biblabel}[1]{\quad#1.}
\makeatother

\date{}

\begin{document}

\begin{flushleft}
{\Large
\textbf{Emergent Behaviors from A Cellular Automaton Model for Invasive Tumor Growth in Heterogeneous Microenvironments}
}
\\
Yang Jiao$^{1\ast}$,
Salvatore Torquato$^{1, 2,\ast\ast}$
\\
\bf{1} Physical Science in Oncology Center, Princeton
Institute for the Science and Technology of Materials, Princeton
University, Princeton New Jersey 08544, USA
\\
\bf{2} Department of Chemistry and Physics,
Princeton Center for Theoretical Science,
Program in Applied and Computational Mathematics,
Princeton University, Princeton New Jersey 08544, USA
\\
$\ast$ E-mail: yjiao@princeton.edu
$\ast\ast$ E-mail: torquato@electron.princeton.edu
\end{flushleft}

\section*{Abstract}
Understanding tumor invasion and metastasis is of crucial importance
for both fundamental cancer research and clinical practice.
\textit{In vitro} experiments have established that the invasive growth
of malignant tumors is characterized by the dendritic invasive branches composed of chains
of tumor cells emanating from the primary tumor mass. The preponderance
of previous tumor simulations focused on non-invasive (or proliferative) growth.
The formation of the invasive cell chains and their interactions
with the primary tumor mass and host microenvironment are not well understood.
Here, we present a novel cellular automaton (CA) model that enables one to efficiently
simulate invasive tumor growth in a heterogeneous host microenvironment.
By taking into account a variety of  microscopic-scale tumor-host interactions,
including the short-range mechanical interactions between
tumor cells and tumor stroma, degradation of extracellular matrix by the
invasive cells and oxygen/nutrient gradient driven cell motions,
our CA model predicts a rich spectrum of growth dynamics and emergent behaviors of invasive tumors.
Besides robustly reproducing the salient features
of dendritic invasive growth, such as least resistance and intrabranch homotype attraction,
we also predict nontrivial coupling of the growth dynamics of
the primary tumor mass and the invasive cells. In addition, we show that
the properties of the host microenvironment can significantly
affect tumor morphology and growth dynamics,
emphasizing the importance of understanding the tumor-host interaction.
The capability of our CA model suggests that well-developed \textit{in silico} tools
could eventually be utilized in clinical situations to predict neoplastic progression
and propose individualized optimal treatment strategies.

\section*{Author Summary}
The goal of the present work is to develop an efficient single-cell based
cellular automaton (CA) model that enables one to investigate the growth dynamics
and morphology of invasive solid tumors. Recent experiments have shown that highly malignant
tumors develop dendritic branches composed of tumor cells that follow each other,
which massively invade into the host microenvironment and ultimately lead
to cancer metastasis. Previous theoretical/computational
cancer modeling neither addressed the question of how such chain-like
invasive branches form nor how they interact with the host microenvironment
and the primary tumor. Our CA model, which incorporates a variety of
microscopic-scale tumor-host interactions (e.g., the mechanical interactions
between tumor cells and tumor stroma, degradation of extracellular matrix by
the tumor cells and oxygen/nutrient gradient driven cell motions), can robustly reproduce
experimentally observed invasive tumor evolution and predict a
wide spectrum of invasive tumor growth dynamics and emergent behaviors in various different heterogeneous
environments. Further refinement of our CA model could eventually lead to the
development of a powerful simulation tool that can be utilized in the clinic
to predict neoplastic progression
and propose individualized optimal treatment strategies.

\section*{Introduction}

Cancer is not a single disease,
but rather a highly complex and heterogeneous set of diseases that can
adapt in an opportunistic manner, even under a variety of stresses.
It is now well accepted that genome level changes in cells, resulting in
the gain of function of oncoproteins or the loss of function of tumor suppressor
proteins, initiate the transformation of normal cells into malignant ones
and neoplastic progression \cite{Co98, hanahan00}. 
In the most aggressive form, malignant cells can leave the primary tumor,
invade into surrounding tissues, find their way into the
circulatory system (through vascular network)
and be deposited at certain organs in the body, leading to the
development of secondary tumors (i.e., metastases) \cite{fearon90}. 

\begin{figure}[bthp]
\begin{center}
$\begin{array}{c@{\hspace{0.6cm}}c}\\
\includegraphics[height=5.25cm,keepaspectratio]{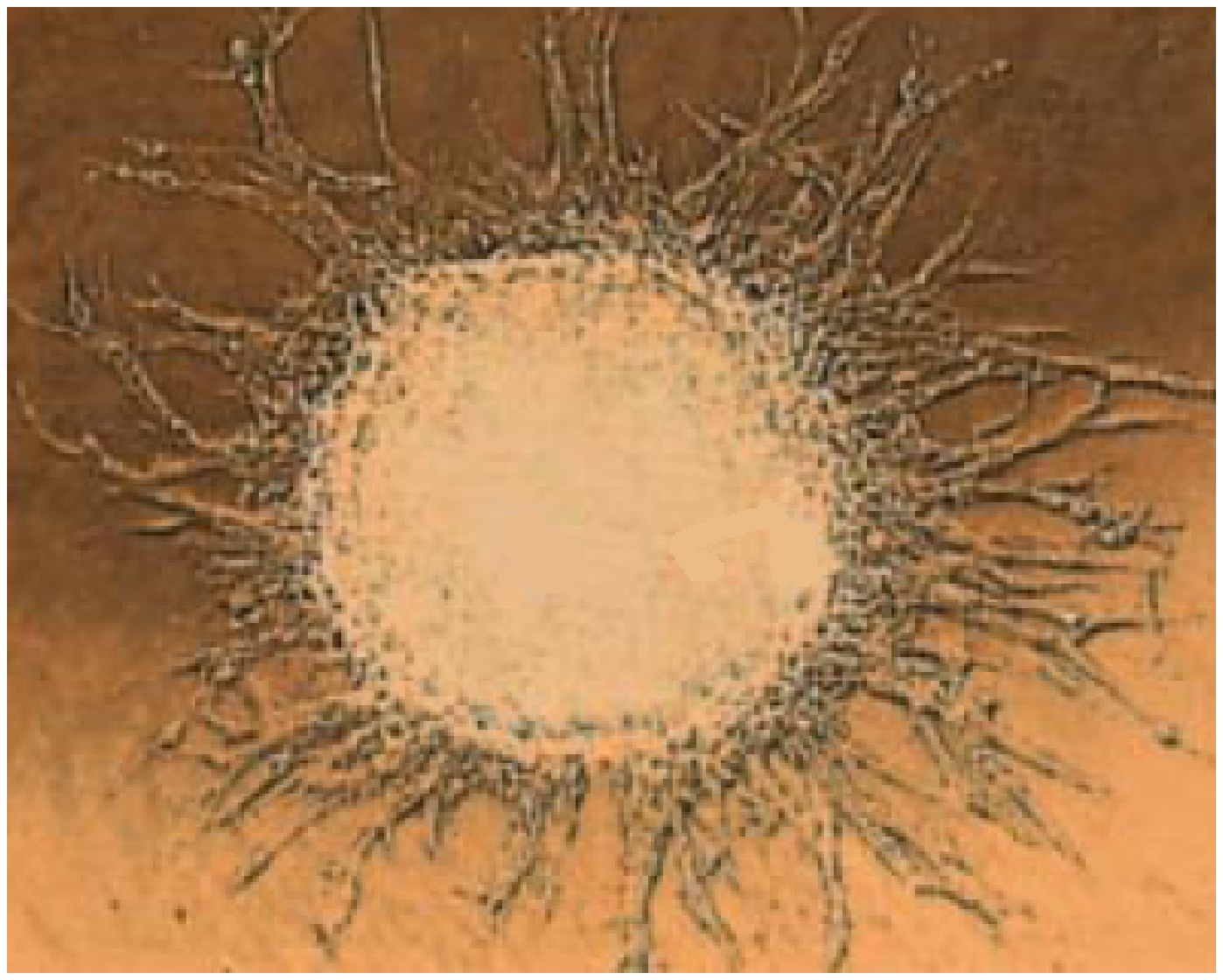} &
\includegraphics[height=5.25cm,keepaspectratio]{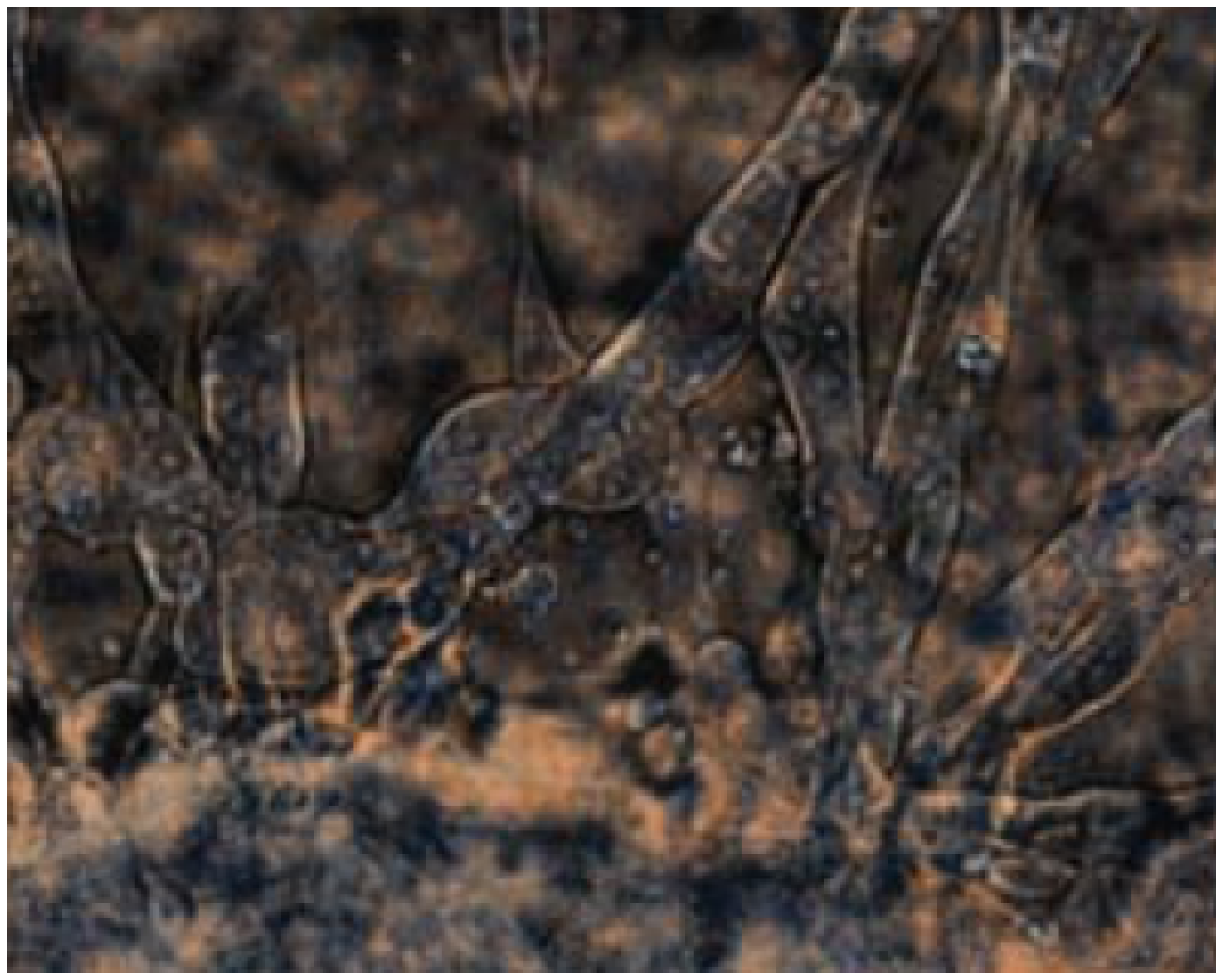} \\
\mbox{\bf (a)} & \mbox{\bf (b)}
\end{array}$
\end{center}
\caption{{\bf GBM multicelluar tumor spheroid (MTS) gel assay showing  dendritic invasive branches.}
(a) The invasive branches centrifugal evolve from the central MTS.
The linear size of central MTS is approximately 400 $\mu$m.
(b) The invasive branches are composed of chains of invasive cells. The
images are adapted from Ref.\cite{deisboeck01}.}
\label{fig_MTS}
\end{figure}

The emergence of invasive behavior in cancer is fatal. For example,
the malignant cells that invade into the surrounding host tissues can quickly
adapt to various environmental stresses and develop resistance to therapies.
The invasive cells that are left behind after resection are responsible for
tumor recurrence and thus an ultimately fatal outcome. Therefore,
significant effort has been expended to understand the mechanisms evolved in the invasive
growth of malignant tumors \cite{hanahan00, deisboeck01, fidler03, kerbel90, liotta03}
and their treatment \cite{gillies04, gatenby09}.
It is generally accepted that the invasive behavior of cancer
is the outcome of many complex interactions occurring between
the tumor cells, and between a tumor and the host
microenvironment \cite{fearon90}. Tumor invasion itself
is a complex multistep process involving homotype detachment,
enzymatic matrix degradation, integrin-mediated heterotype adhesion,
as well as active, directed and random motility \cite{deisboeck01}.
In recent \textit{in vitro} experiments involving glioblastoma multiforme (GBM),
the most malignant brain cancer, it has been observed that
dendritic invading branches composed of chains
of tumor cells are emanating from the primary tumor mass; see Fig.~\ref{fig_MTS}.
Such invasive behaviors are characterized by intrabranch homotype
attraction and least resistance \cite{deisboeck01}.

Although recently progress has been made in understanding certain
aspects of the complex tumor-host interactions that may be responsible
for invasive cancer behaviors \cite{deisboeck01, fred91, brand00, kitano04},
many mechanisms are either not fully understood or are unknown at the moment.
Even if all of the mechanisms for cancer invasion could be identified,
it is still not clear that progress in understanding neoplastic progression
and proposing individualized optimal treatment strategies could be made
without the knowledge of how these different mechanisms couple
to one another and to the heterogeneous host microenvironment in which tumor grows \cite{To11}.
Theoretical/computational cancer modeling that integrates apart
mechanisms, when appropriately linked with experimental
and clinical data, offers a promising avenue for a better understanding
of tumor growth, invasion and metastasis.
A successful model would enable one to broaden the conclusions drawn
from existing medical data, suggest new experiments, test hypotheses,
predict behavior in experimentally unobservable situations, and be
employed for early detection and prognosis.

Indeed, cancer modeling has been a very active area of research for
the last two decades (see Refs.~\cite{Byrne10} and \cite{To11} for recent reviews).
A variety of interactions between tumor cells and between tumor and its
host microenvironment have been investigated \cite{chaplain98, kansal00a, kansal00b, kansal02,
gevertz06, jana08, jana09, anderson05, anderson06,
bankhead07, gatenby96, gatenby96b, gatenby06, bellomo00, scalerandi01, scalerandi02, kim10},
via continuum \cite{gatenby96, gatenby96b, gatenby06, gatenby06b, kim10}, discrete \cite{kansal00a, jana08, stein07}
or hybrid \cite{gevertz06, jana09, anderson05, anderson06} mathematical models.
Very recently, multiscale mathematical models \cite{anderson05, anderson06, gatenby06b}
have been employed to study the effects of
the host microenvironment on the morphology and phenotypic evolution of invasive tumors
and it has been shown that microenvironmental heterogeneity can dramatically
affect the growth dynamics of invasive tumors. Although the simulated
tumors showed certain invasive characteristics (e.g., developing protruding
surfaces), no dendritic invasive branches emerged.

In response to the challenge to develop an ``Ising'' model for
cancer growth \cite{To11}, we generalize a cell-based discrete
cellular automaton (CA) model that we have developed
\cite{kansal00a, kansal00b, kansal02, jana08, jana09} to
investigate the invasive growth of malignant tumors in
heterogeneous host microenvironments. To the best of our knowledge,
this generalized CA model is the first to investigate the
formation of invasive cell chains and their interactions with the
primary tumor mass and the host microenvironment. Our cellular
automaton model takes into account a variety of  microscopic-scale
tumor-host interactions, including the short-range mechanical
interactions between tumor cells and tumor stroma, the degradation
of extracellular matrix by the invasive cells and oxygen/nutrient
gradient driven cell motions and thus, it can predict a wide range
of growth dynamics and emergent behaviors of invasive tumors. In
particular, our CA model robustly reproduces the salient features
of dendritic invasive growth observed in experiments, which is
characterized by least resistance and intrabranch homotype
attraction. The model also predicts nontrivial coupling of the
growth dynamics of the primary tumor mass and the invasive cells,
e.g., the invasive cells can facilitate the growth of primary
tumor in harsh microenvironment. Moreover, we show that the
properties of the host microenvironment can significantly affect
tumor growth dynamics and lead to a variety of tumor morphology.
These emergent behaviors naturally arise due to various
microscopic-scale tumor-host interactions, which emphasizes the
importance of taking into account microenvironmental heterogeneity
in understanding cancer. Further refinement of our model could
eventually lead to the development of a powerful \textit{in
silico} tool that can be utilized in the clinic. As a
demonstration of the capability and versatility of our CA model,
we mainly consider invasive tumor growth in two dimensions,
although the model is easily extended to three dimensions. Indeed,
the algorithmic details of the model are given for any spatial
dimension.

\section*{Materials and Methods}

\subsection*{Biophysical Background of the CA Model}


\subsubsection*{Voronoi Tessellation: The Underlying Cellular Structure}

\begin{figure}[bthp]
\begin{center}
$\begin{array}{c@{\hspace{1.25cm}}c}\\
\includegraphics[height=5.0cm,keepaspectratio]{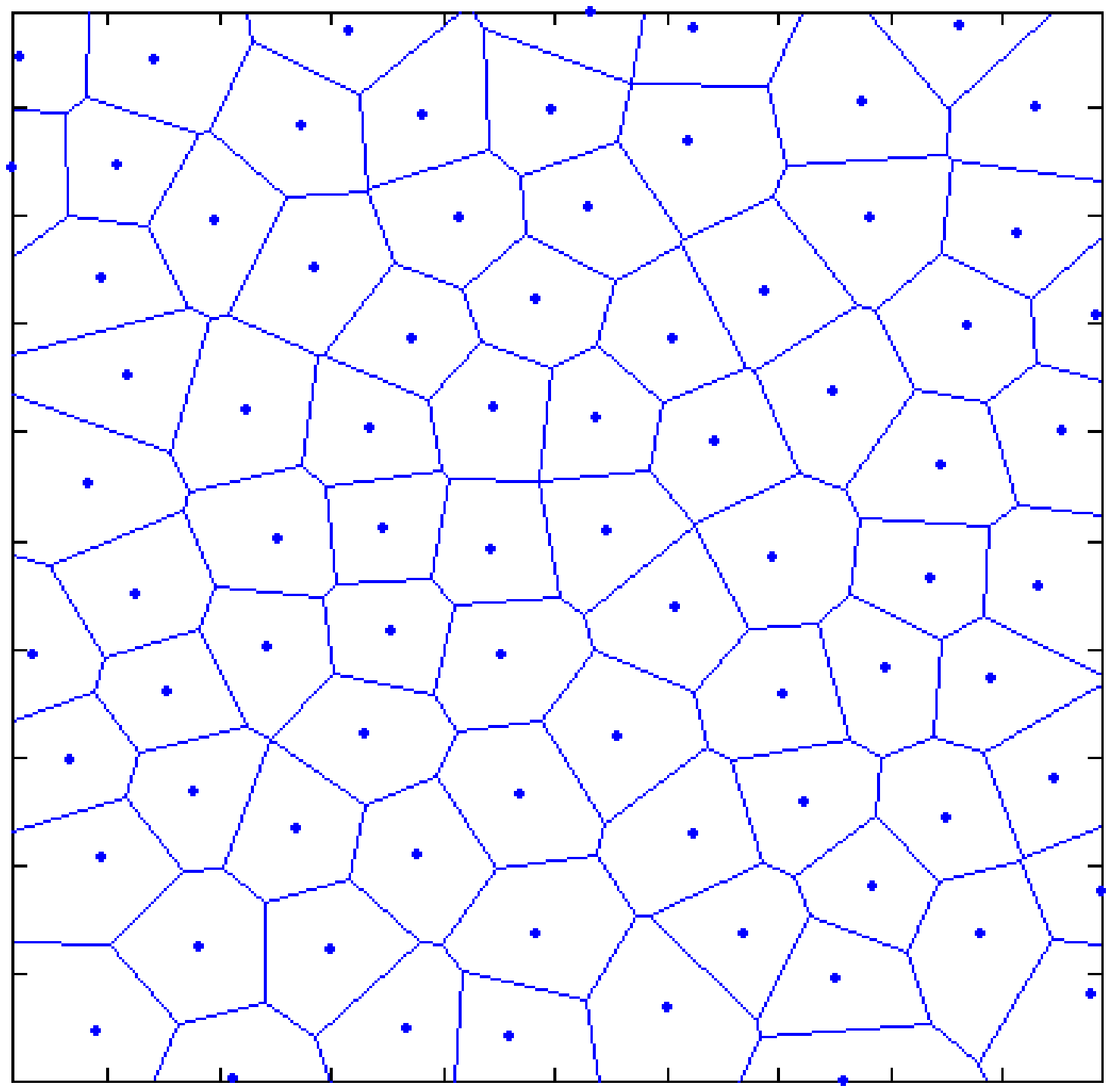} &
\includegraphics[height=5.3cm,keepaspectratio]{fig2b.eps} \\
\mbox{\bf (a)} & \mbox{\bf (b)}
\end{array}$
\end{center}
\caption{{\bf A 2D Voronoi tessellation and the associated point configuration.}
(a) A Voronoi tessellation of the 2D plane into polygons which are the
automaton cells in our model.
(b) The associated point configuration for the tessellation, generated by
randomly placing nonoverlap circular disks in a prescribed region, i.e.,
the random sequential addition process \cite{torquato}.}
\label{fig_Vor}
\end{figure}

The underlying cellular structure is modeled using a
Voronoi tessellation of the space into polyhedra \cite{torquato},
based on centers of spheres in a packing generated by a random sequential
addition (RSA) process \cite{kansal00a, jana09} (see Fig.~\ref{fig_Vor}).
In particular, nonoverlapping spheres are randomly and sequentially
placed in a prescribed region until there is no void space for additional
spheres, i.e., saturation is achieved. Such a saturated RSA packing possesses
relative small variations in its Voronoi polyhedra and thus, has
served as models for many biological systems \cite{To_RMP, Patel09}.
We refer to the polyhedra associated with the Voronoi
tessellation as automaton cells. These automaton cells may
correspond to real cells or tumor stroma (e.g., clusters of the ECM macromolecules). In previous studies, such
automaton cells have represented clusters of real cells of
various sizes or have implicitly represented healthy tissues \cite{kansal00a}.
Thus, the Voronoi tessellation associated with RSA sphere centers provides a
highly flexible model for real-cell aggregates with a high degree of shape isotropy.
For example, one can use a variable automaton cell size to
simulate avascular tumor growth from a few malignant cells
to its macroscopic size \cite{kansal00a}. In addition, such a Voronoi tessellation
can reduce the undesired growth bias due to the anisotropy
of ordered tessellations based on square and simple cubic lattices.

Since our new CA model explicitly takes into account the
interactions between a single cell and its neighbors and
microenvironment, each automaton cell here represents either a
single tumor cell or a region of tumor stroma. Thus, the linear
size of a single automaton cell is approximately $15 - 20~\mu$m
and the linear size of the 2D simulation domain is approximately 5
mm, which contains $\sim 100~000$ automaton cells. In the current
model, we mainly focus on the effects of ECM
macromolecule density and ECM degradation by malignant cells on tumor growth.
Henceforth, we will refer to the host microenvironment (or tumor
stroma) as ``ECM'' for simplicity. Each ECM associated
automaton cell is assigned a particular density
$\rho_{\mbox{\tiny{ECM}}}$, representing the density of the ECM
molecules within the automaton cell. A tumor cell can occupy an
ECM associated automaton cell only if the density of this
automaton cell $\rho_{\mbox{\tiny{ECM}}} = 0$, which means that
either the ECM is degraded or it is deformed (pushed away) by the
proliferating tumor cells.

\subsubsection*{Microenvironment Heterogeneity}

The microenvironment in which tumor grows is usually highly heterogeneous, composed of various types of
stromal cells and ECM structures. The ECM is a complex mixture of
macromolecules that provide mechanical supports for the tissue (such as collagens) and
those that play an important role for cell adhesion and motility (such as laminin and fibronectin) \cite{anderson05, jana_brm, ECM}.
For different individuals with tumors, the ECM in
the host microenvironments generally possess distinct mechanical and transport properties.
By explicitly representing the ECM using automaton cells with different macromolecule densities,
the effects of microenvironment heterogeneity
on tumor growth can be very well explored. For example,
various distributions of the ECM densities (i.e., the densities of the
ECM macromolecules) can be employed to mimic
the actual heterogeneous host microenvironment of the tumor. Certain
tumor stroma contains fibroblasts, which actively produce ECM macromolecules
leading to a higher ECM density around these cells.
The automaton cells representing the ECM
with larger densities are considered to be more rigid and more difficult
to degrade. Since each automaton cell associated with the ECM has its own density, this
allows a variation of ECM characteristics on the length scale comparable
to that of a single tumor cell.

In addition, the tumor in our model is only allowed to grow in a compact
growth-permitting region. This is to mimic the physical confinement of the host microenvironment,
such as the boundary of an organ. In other words, only automaton
cells within this region can be occupied by the cells of the tumor as it grows.
In general, the growth-permitting
region can be of any shape that best models the organ shape. Here
we simply choose a spherical region to study the effects of the heterogeneous
ECM on tumor growth. More sophisticated growth-region shapes have
been employed to investigate the effects of physical confinement on tumor growth \cite{jana08, jana09}.
Furthermore, we assume a constant radially symmetric nutrient/oxygen
gradient in the growth-permitting region with the highest nutrition
concentration at the boundary of this region, i.e., it is a vascular boundary.
However, this assumption can also be relaxed.

\subsubsection*{Tumor Cell Phenotypes and Interactions with the Host Microenvironment}

For highly malignant tumors, we consider the cells to be of one of
the two classes of phenotypes: either invasive or non-invasive.
Following Ref.\cite{kansal00a}, the non-invasive
cells remain in the primary tumor and can be proliferative, quiescent
or necrotic, depending on the nutrition supply they get. For avascular
tumor growth, as we focus here, the nutrition the tumor cells can get
are essentially those diffuse into the tumor through its surface. As
the tumor grows, the amount of nutrition supply,
which is proportional to the surface area of the tumor, cannot
meet the needs of all cells whose number increases with the tumor volume,
leading to the development of necrotic and quiescent regions.
Following Ref.\cite{kansal00a}, characteristic diffusion lengths are employed to
determine the states of a non-invasive cell. For example,
quiescent cells more than $\delta_n$ away from
the tumor surface become necrotic (see details in the next section).
The diffusion length $\delta_n$ (also the characteristic thickness of
the rim of living tumor cells) depends on
the size of the primary tumor.

As a proliferative cell divides, its daughter cell effectively
pushes away/degrades the surrounding ECM and occupies the
automaton cell originally associated with the ECM \cite{mech1,
mech2, degrad1}. It is easier for a tumor cell to take up an ECM
associated automaton cell with lower density (i.e., less rigid ECM
regions) than that with higher density (i.e., more rigid ECM
regions) and thus, the tumor growth is affected by the ECM
heterogeneity through the local mechanical interaction between
tumor cells and the ECM. If there is no space available for the
placement of a daughter cell within a distance $\delta_p$ from the
proliferative cell, the proliferative cell turns quiescent.

The invasive cells are considered to be mutant daughters of the
proliferative cells \cite{mutant1}, which can gain a variety of degrees of ECM
degradation ability $\chi$ (i.e., the matrix-degradative enzymes)
and motility $\mu$ that allow them to leave the
primary tumor and invade into surrounding microenvironment \cite{degrad2}.
We consider the invasive cells can move from one automaton cell
to another only if the ECM in the target automaton cell
is completely degraded (i.e., with $\rho_{\mbox{\tiny{ECM}}} = 0$). Each trial
move of an invasive cell involves the degradation of the ECM in its neighbor
automaton cells, followed by a possible move to one of the automaton cells whose ECM
is completely degraded; otherwise the invasive cell does not move.
The number of trial moves of an invasive cell and to what extent it
degrades the ECM are respectively determined by $\mu$ and $\chi$
(see the following section for details).
The oxygen/nutrient gradient also drives the invasive cells to move as far as
possible from the primary tumor \cite{mot}, which takes up the majority of oxygen/nutrients.
The motility $\mu$ is the maximum possible number of such trial moves.
In addition, we assume that the invasive cells do not divide as they migrate.


\subsection*{Algorithmic Details}



We now provide specific details for the CA model to study invasive tumor growth in
confined heterogeneous microenvironment. In what follows, we will simply refer to the
primary tumor as ``the tumor'' and explicitly use ``invasive'' when considering invasive cells.
After generating the automaton cells by Voronoi tessellation of RSA sphere centers,
an ECM macromolecule density $\rho_{\mbox{\tiny{ECM}}} \in (0, ~1)$ is assigned to
each automaton cell within the growth-permitting region, which represents
the heterogeneous host microenvironment. Then a tumor is introduced by designating any
one or more of the automaton cells as proliferative
cancer cells. Time is then discretized into units that represent one real
day. At each time step:

\begin{itemize}
\item Each automaton cell is checked for type: invasive, proliferative, quiescent,
necrotic or ECM associated. Invasive cells degrade and migrate into the ECM surrounding the
tumor. Proliferative cells are actively dividing cancer cells, quiescent
cancer cells are those that are alive, but do not have enough oxygen and
nutrients to support cellular division and necrotic cells are dead cancer cells.

\item All ECM associated automaton cells and tumorous necrotic cells are inert
(i.e., they do not change type).

\item Quiescent cells more than a certain distance $\delta_n$ from the tumor's
edge are turned necrotic. The tumor's edge, which is assumed to be the
source of oxygen and nutrients, consists of all ECM associated automaton cells that border
the neoplasm. The critical distance $\delta_n$ for quiescent cells to turn
necrotic is computed as follows:
\begin{equation*}
\label{eq_delta}
\delta_n = aL_t^{(d-1)/d},
\end{equation*}
where $a$ is prescribed parameter (see Table \ref{tab_Param}), $d$ is the spatial dimension and
$L_t$ is the distance between the
geometric centroid ${\bf x}_c$ of the tumor and the tumor edge cell that is closest to the
quiescent cell under consideration. The position of the tumor centroid ${\bf x}_c$ is given by
\begin{equation*}
\label{eq_xc}
{\bf x}_c = \frac{{\bf x}_1 + {\bf x}_2 + \cdots + {\bf x}_N}{N},
\end{equation*}
where $N$ is the total number of noninvasive cells contained in the tumor, which
is updated when a new noninvasive daughter cell is added to the tumor.

\item Each proliferative cell will attempt to divide with probability $p_{div}$
into the surrounding ECM (i.e., the automaton cells associated with the ECM) by
degrading and pushing away the ECM in that automaton cell.
We consider that $p_{div}$ depends on both the physical confinement imposed by the
boundary of the growth-permitting region and the local mechanical interaction
between the tumor cells and the ECM, i.e.,
\begin{equation*}
\label{eq_pdiv}
p_{div} = \left\{
\begin{array}{ll}
 & \mbox{if any ECM associated automaton cell within} \\
\frac{p_0}{2}[(1-\frac{r}{L_{max}})+(1-\rho_{\mbox{\tiny{ECM}}})] & \mbox{the predefined growth distance is in the growth-}\\
 & \mbox{permitting microenvironment} \\  \\
 & \mbox{if no ECM associated automaton cell within} \\
0 & \mbox{the predefined growth distance is in the growth-}\\
& \mbox{permitting microenvironment}
\end{array}
\right.
\end{equation*}
where $p_0$ is the base probability of division (see Table \ref{tab_Param}),
$r$ is the distance of the dividing cell from the tumor centroid,
$L_{max}$ is the distance between the closest growth-permitting
boundary cell in the direction of tumor growth and the tumor's
geometric centroid ${\bf x}_c$ and $\rho_{\mbox{\tiny{ECM}}}$ is the ECM density of the
automaton cell to be taken by the new tumor cell. When a
ECM associated automaton cell is taken by a tumor cell, it density is set to be zero.
The predefined growth distance ($\delta_p$) is described in a following bullet point.

\item If a proliferative cell divides, it can produce a mutant daughter cell
possessing an invasive phenotype with a prescribed probability $\gamma$ (i.e., the
mutation rate). The invasive daughter cell gains ECM degradation ability
$\chi$ and motility $\mu$, which enables it to leave the
tumor and invade into surrounding ECM. The rules for updating invasive
cells are given in a following bullet point. If the daughter cell is noninvasive,
it is designated as a new proliferative cell.

\item A proliferative cell turns quiescent if there is no space available for
the placement of a daughter cell within a distance $\delta_p$ from the
proliferative cell, which is given by
\begin{equation*}
\label{eq_deltap}
\delta_p = bL_t^{(d-1)/d},
\end{equation*}
where $b$ a nutritional parameter (see Table \ref{tab_Param}), $d$ is the
spatial dimension and $L_t$ is the
distance between the geometric tumor centroid ${\bf x}_c$ and the tumor edge cell that is
closest to the proliferative cell under consideration.

\item An invasive cell degrades the surrounding ECM (i.e., those in the neighboring
automaton cells of the invasive cell) and can move from
one automaton cell to another if the associated ECM is completely degraded.
For an invasive cell with motility $\mu$ and ECM degradation ability $\chi$,
it will make $m$ attempts to degrade the ECM in the neighboring automaton cells
and jump to these automaton cells, where $m$ is an arbitrary integer in $[0,~\mu]$.
For each attempt, the surrounding ECM density $\rho_{\mbox{\tiny{ECM}}}$ is
decreased by $\delta\rho$, where $\delta\rho$ is an arbitrary number in $[0,~\chi]$.
Using random numbers for ECM degradation ability and cellular motility
is to take into account tumor genome heterogeneity, which is manifested as
heterogeneous phenotypes (such as different $m$ and $\delta\rho$).
When the ECM in multiple neighboring automaton cells of
the invasive cell are completely degraded (i.e., $\rho_{\mbox{\tiny{ECM}}} = 0$), the invasive cell moves
in a direction that maximizes the nutrients
and oxygen supply. Here we assume that the migrating invasive cells do not
divide. The degraded ECM shows the invasive path of the tumor.

\end{itemize}

\begin{figure}[bthp]
\begin{center}
$\begin{array}{c@{\hspace{0.6cm}}c@{\hspace{0.6cm}}c@{\hspace{0.6cm}}c}\\
\includegraphics[height=3.25cm,keepaspectratio]{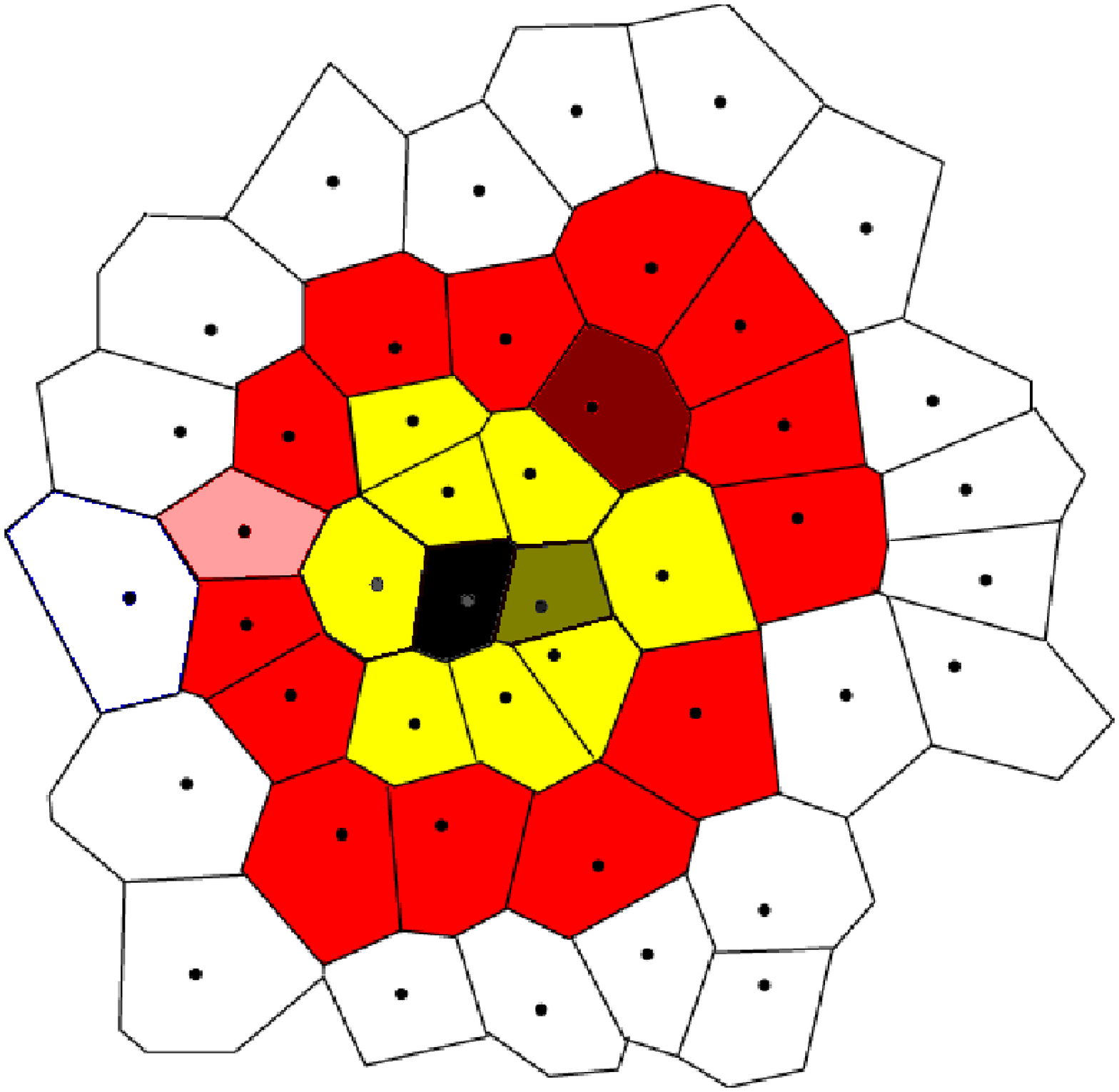} &
\includegraphics[height=3.25cm,keepaspectratio]{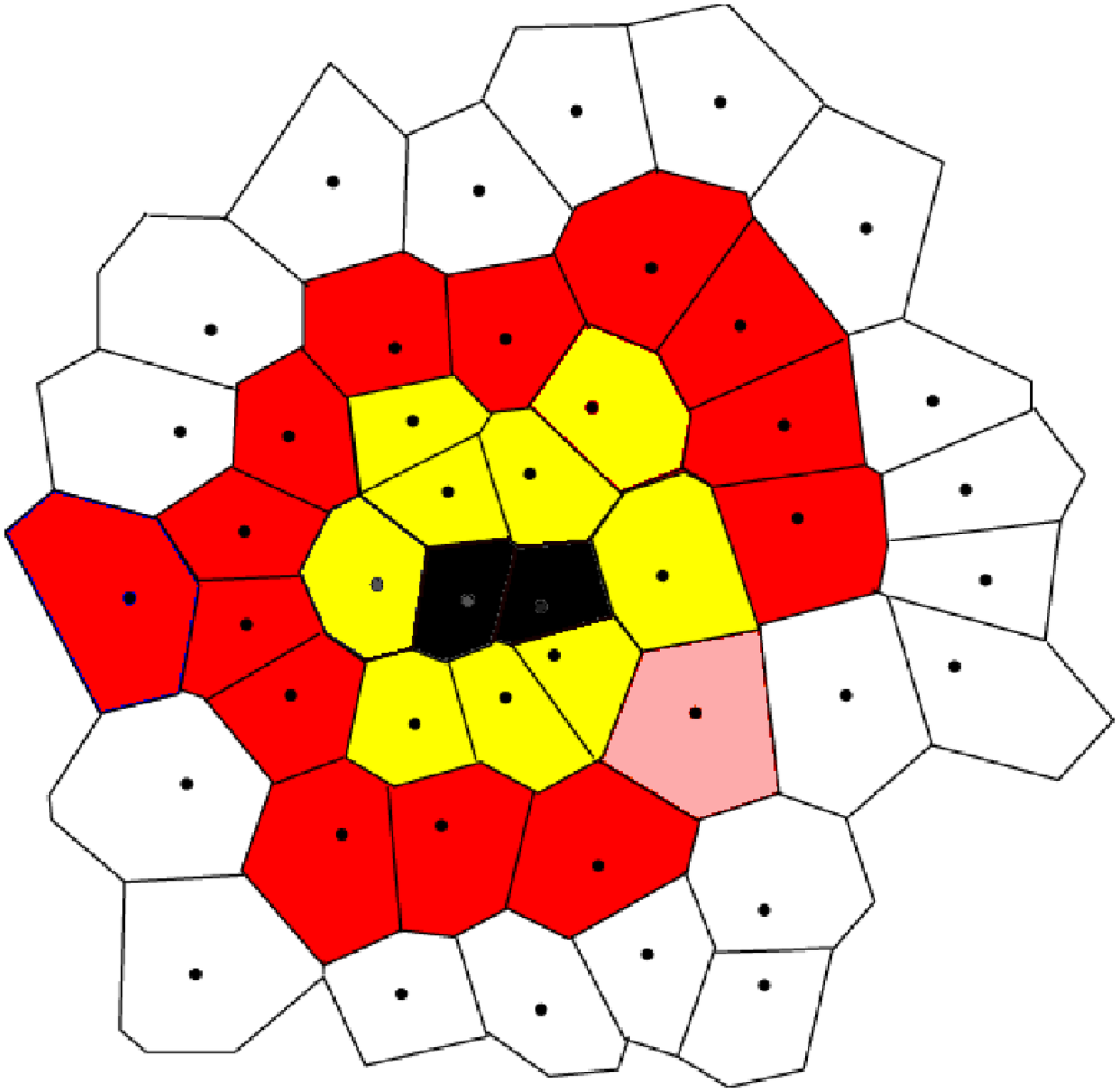} &
\includegraphics[height=3.25cm,keepaspectratio]{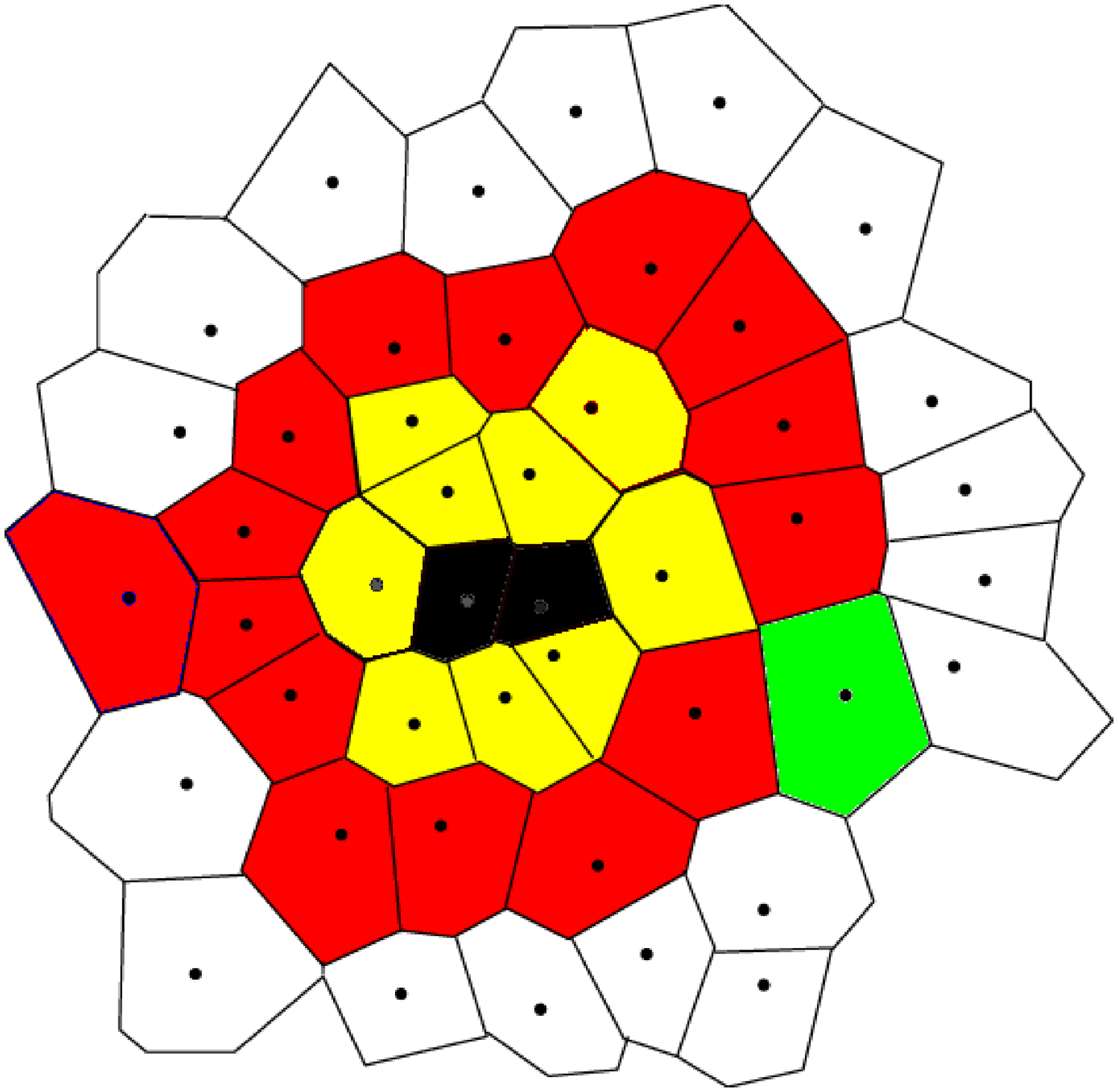} &
\includegraphics[height=3.25cm,keepaspectratio]{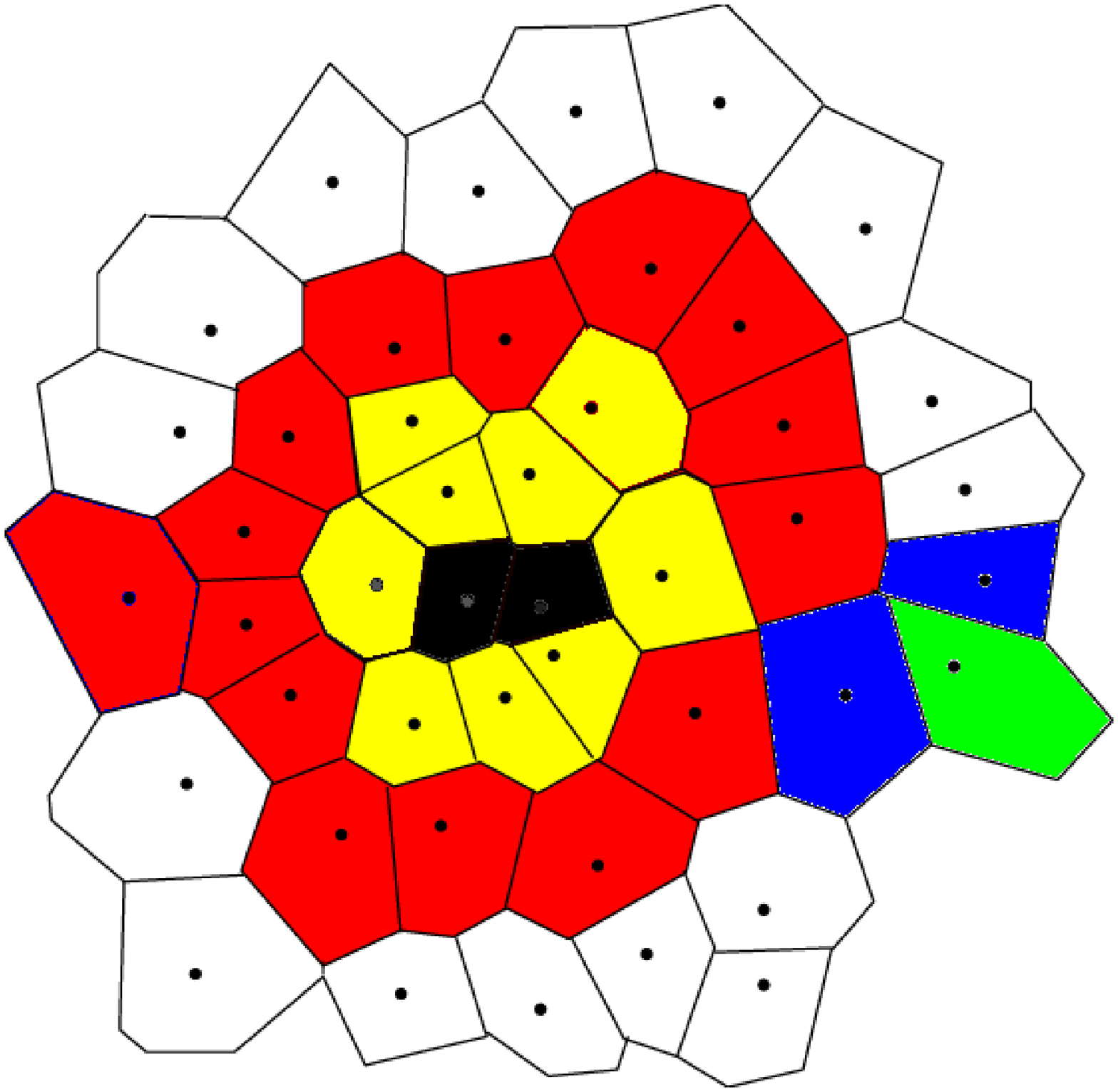} \\
\mbox{\bf (a)} & \mbox{\bf (b)} & \mbox{\bf (c)} & \mbox{\bf (d)}
\end{array}$
\end{center}
\caption{{\bf Illustration of cellular automaton rules.}
Necrotic cells are black, quiescent cells are yellow, proliferative cells are red
and invasive tumor cells are green. The ECM associated automaton cells are white and the degraded ECM is blue.
(a) A proliferative cell (dark red) is too far away from the tumor edge to
get sufficient nutrients/oxygen and it will turn quiescent in panel (b).
A quiescent cell (dark yellow) is too far away from the tumor edge and it will turn necrotic in panel (b).
Another proliferative cell (light red) will produce a daughter proliferative cell in panel (b).
(b) The dark red proliferative cell and the dark yellow quiescent cell in panel (a)
turned quiescent and necrotic, respectively. The light red proliferative cell in panel (a)
produced a daughter cell. Another proliferative cell (light red) will produce a mutant
invasive daughter cell. (c) The light red proliferative cell in (b) produced
an invasive cell. (d) The invasive cell degraded the surrounding ECM and moved
to another automaton cell.}
\label{fig_CA}
\end{figure}

\noindent The aforementioned automaton rules are briefly illustrated in
Fig.~\ref{fig_CA}. We note that non-invasive tumor growth can be studied by
imposing a mutation rate $\gamma = 0$. This enables us to compare the growth dynamics
of invasive and non-invasive tumors and in turn to investigate
the effects of coupling growth of the primary tumor mass and the invasive cells.
Although we only consider spherical-growth-permitting regions here,
the CA rules given above allow growth-permitting regions with arbitrary shapes.
The important parameters mentioned in the bullet points above are summarized in Table \ref{tab_Param}.
In the following, we will employ our CA model to investigate the growth
dynamics of malignant tumors with different degrees of invasiveness
in variety of different heterogeneous microenvironments.

\subsection*{Quantitative Metrics for Tumor Morphology}


To characterize quantitatively the morphology of simulated tumors,
we present several scalar metrics that capture the salient
geometric features of the primary tumor, dendritic invasive
branches or the entire invasive pattern. These metrics include the
ratio $\beta$ of invasive area over tumor area (defined below),
the specific surface $s$ of the invasive pattern, the asphericity
$\alpha$ of the primary tumor and the angular anisotropy metric
$\psi$ for the invasive branches. The metrics are computed for all
simulated tumors and compared to available experimental data. We
note that the invasive pattern associated with a neoplasm includes
both the primary tumor and the invasive branches.

\begin{figure}[bthp]
\begin{center}
$\begin{array}{c@{\hspace{0.6cm}}c@{\hspace{0.6cm}}c}\\
\includegraphics[height=4.25cm,keepaspectratio]{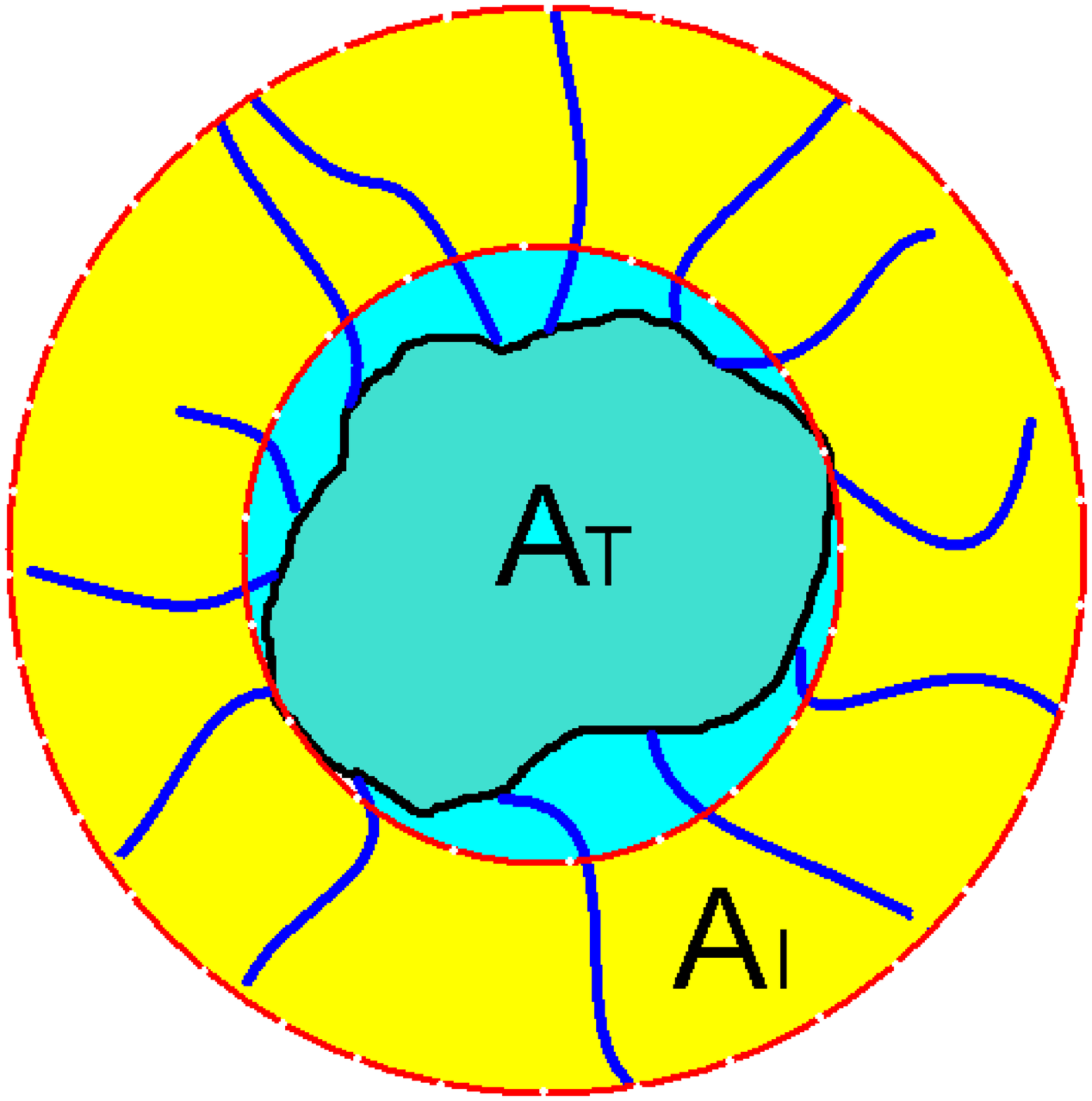} &
\includegraphics[height=4.25cm,keepaspectratio]{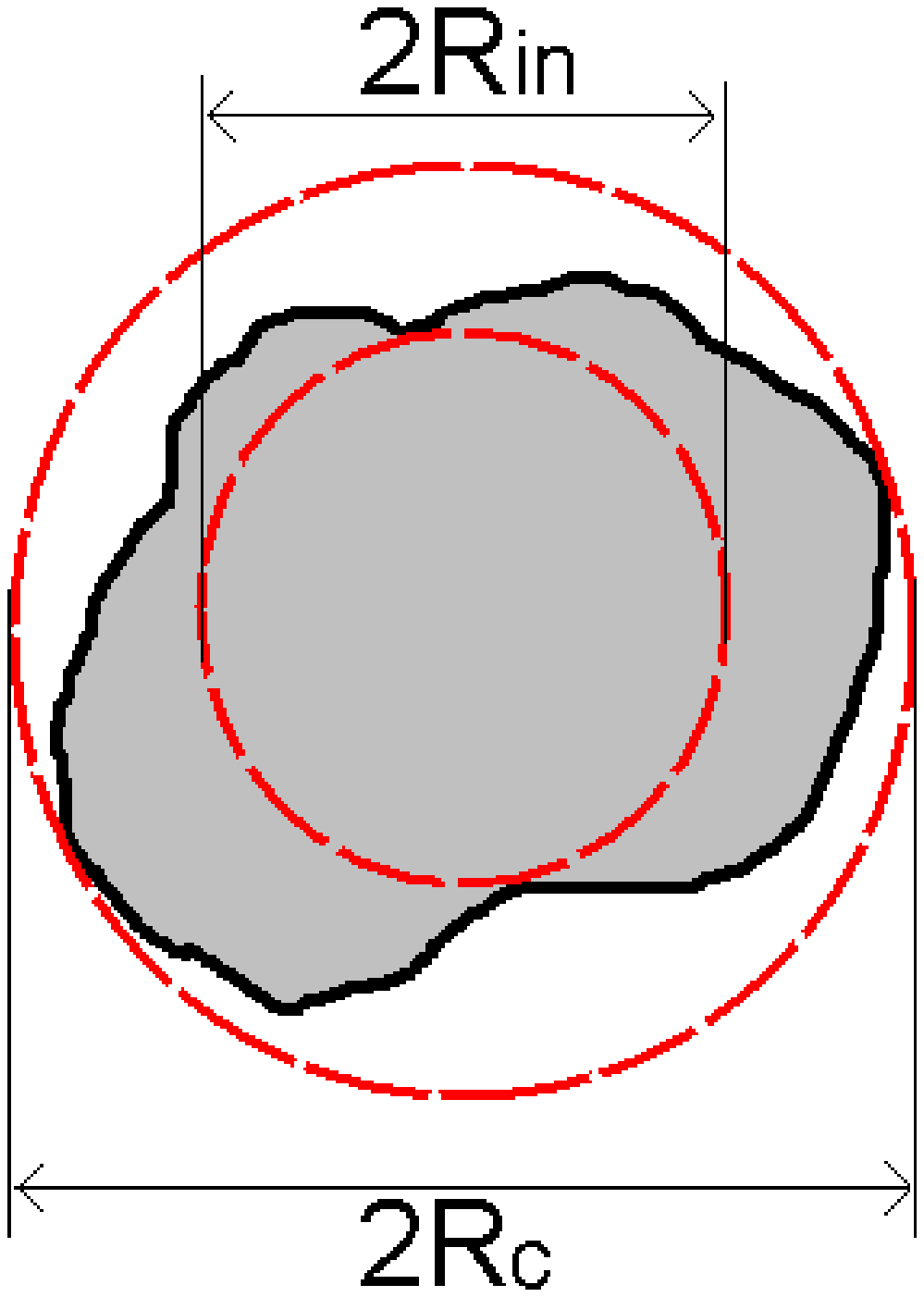} &
\includegraphics[height=4.25cm,keepaspectratio]{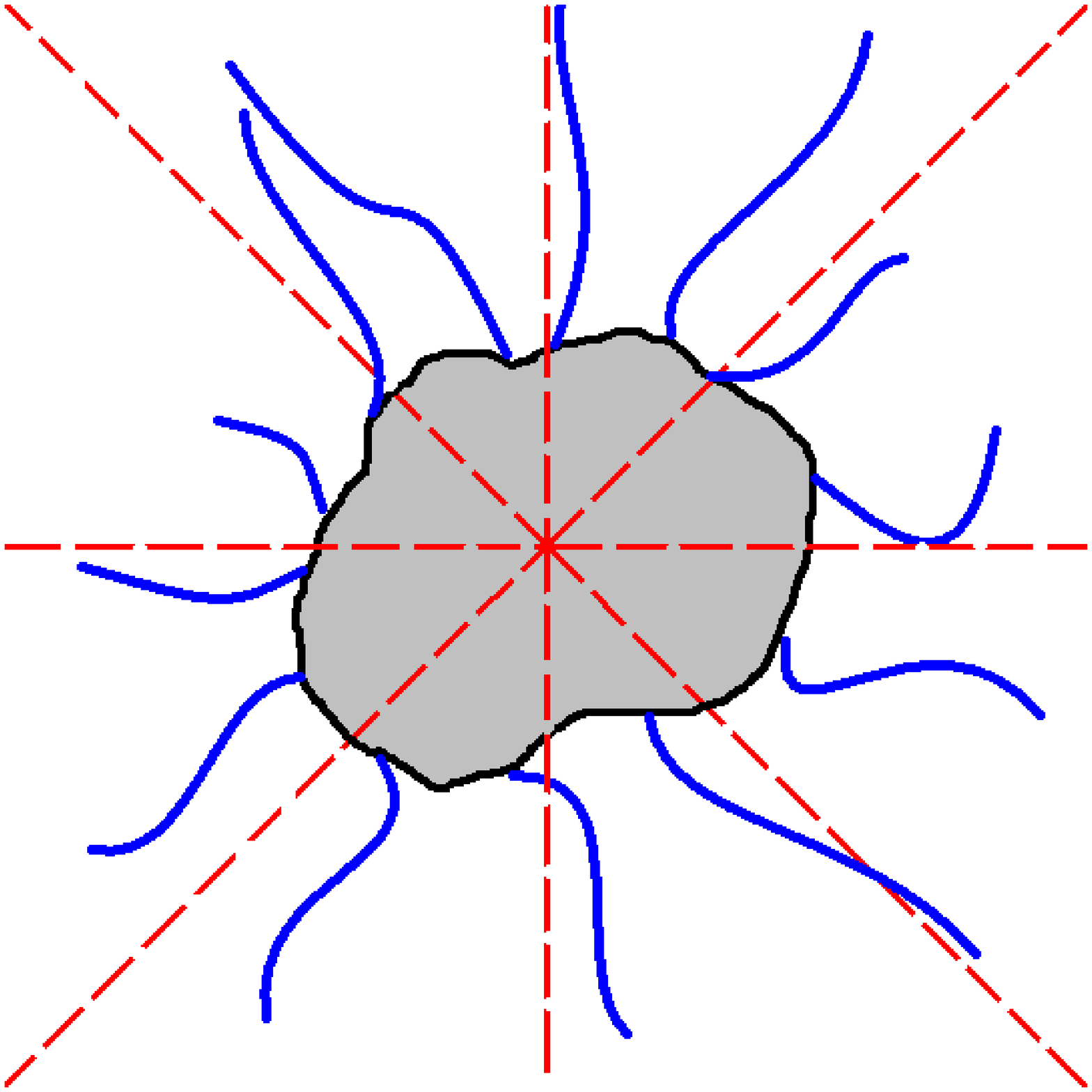} \\
\mbox{\bf (a)} & \mbox{\bf (b)} & \mbox{\bf (c)}
\end{array}$
\end{center}
\caption{{\bf Schematic illustration of the quantities in the definitions of tumor morphology metrics.}
(a) Invasive area $A_I$ and tumor area $A_T$ associated with the invasive pattern.
(b) Circumcircle with radius $R_c$ and incircle with radius $R_{in}$ associated with the primary tumor.
(c) Evenly dividing the invasive pattern into $n_a = 8$ sectors for computing angular
anisotropy metric $\psi$.}
\label{Metric}
\end{figure}

Following Ref.~\cite{deisboeck01}, the tumor area $A_T$ is defined as
the area of the circumcircle of the primary tumor (see Fig.~\ref{Metric}(a))
and the invasive area $A_I$ is the area of the region between the effective circumcircle
of the invasive pattern and the circumcircle of the primary tumor (see Fig.~\ref{Metric}(a)).
The radius of the effective circumcircle of the invasive pattern is
defined to be the average distance from the invasive branch tip to the tumor center.
The ratio $\beta = A_I/A_T$ as a function of time $t$
reflects the degree of coupling between the primary tumor and the invasive cells.
If $\beta(t)$ is linear in $t$, there is no coupling; otherwise the two are coupled.

The specific surface $s$ \cite{torquato} for the invasive pattern is defined as the ratio of
the total length of the perimeter of the invasive pattern over its total area.
In general, $s$ is inversely proportional to the size of the tumor and thus, large
tumors have small $s$ values. Moreover, given the tumor size, tumors with a large number of
long dendritic invasive branches possess a large value of $s$. And $s$ is minimized
for perfectly circular tumors with $s=2/R_T$, where $R_T$ is the radius.
Since $s$ depends on the size of the tumor, which makes it difficult to
compare tumors with different sizes, in the calculations that follow we employ a normalized $s$ with respect to $2/R_T$
for an arbitrary-shaped tumor with effective radius $R_T$ (i.e., the average distance from tumor edge to tumor center).
For simplicity, we will still refer to the normalized specific surface as
``specific surface'' and designate it with symbol $s$.

The asphericity $\alpha$ of the primary tumor is defined as the ratio
of the radius of circumcircle $R_c$ of the primary tumor over its incircle
radius $R_{in}$ \cite{ToJi09}, i.e., $\alpha = R_c/R_{in}$ (see Fig.~\ref{Metric}(b)).
A large $\alpha$ value indicates a large deviation of the shape of
primary tumor from that of a perfect circle, i.e., the tumor is more anisotropic.

To quantify the degree of anisotropy of the invasive branches, we
introduce the angular anisotropy metric $\psi$. In particular, the
entire invasive pattern is evenly divided into $n_a$ sectors with lines
emanating from the tumor center (see  Fig.~\ref{Metric}(c)).
The angular anisotropy metric $\psi$ is defined as
\begin{equation}
\label{eq_psi}
\psi = \frac{\sum_{i=1}^{n_a}|\overline{\ell}(i)-\ell _{ave}|}{n_a \cdot \ell_{ave}},
\end{equation}
where $\overline{\ell}(i)$ is the average length of the invasive branches
within the $i$th sector and
\begin{equation}
\ell_{ave} = \frac{\sum_{i=1}^{n_a}\overline{\ell}(i)}{n_a},
\end{equation}
is the average length of all invasive branches. For tumors with
invasive branches of similar lengths that are uniformly angularly
distributed, the metric $\psi$ is small. Large fluctuations of
both invasive branch length and angular distribution can lead to
large $\psi$ values. In the following, we use $n_a = 16$ to
compute $\psi$ for the simulated invasive tumors.

\section*{Results}

\subsection*{Model Validation}

To verify the robustness and predictive capacity of our CA model,
we first employ it to quantitatively reproduce the observed invasive
growth of a GBM multicellular tumor spheroid (MTS) in vitro \cite{deisboeck01}.
In particular, the boundary of the growth-permitting region is considered to be
vascularized, i.e., a growing tumor can receive oxygen and nutrients
from the growth-permitting region. A constant radially symmetric nutrient/oxygen
gradient in the growth-permitting region with the highest nutrient/oxygen
concentration at the vascular boundary is used. Initially, approximately 250 proliferative
tumor cells are introduced at the center of the growth-permitting
region with homogeneous ECM and tumor growth is started.
This corresponds to an initial MTS with diameter $D_{\mbox{\tiny MTS}} \approx 310~ \mu$m
which is consistent with the in vitro experiment set-up \cite{deisboeck01}.
The following values of the growth and invasiveness parameters are used: $p_0 = 0.384$,
$a = 0.58$ mm$^{1/2}$, $b = 0.30$ mm$^{1/2}$, $\gamma = 0.05$, $\chi = 0.55$, $\mu = 3$.
Note that the value of $p_0$ corresponds to a cell doubling time of 40 hours, which
is consistent with the reported experimental data \cite{deisboeck01}.
 A small value of ECM density $\rho_{\mbox{\tiny ECM}} = 0.15$
is used, which corresponds to the soft DMEM medium used in the experiment \cite{deisboeck01}.
In the visualizations of the tumor that follow, we use the following convention:
the ECM in the growth-permitting region is white, and gray outside this region.
The ECM degraded by the tumor cells is blue.
In the primary tumor, necrotic cells are black, quiescent cells are yellow
and proliferative cells are red. The invasive tumor cells are green.

\begin{figure}[bthp]
\begin{center}
$\begin{array}{c@{\hspace{0.6cm}}c@{\hspace{0.6cm}}c}\\
\includegraphics[height=4.25cm,keepaspectratio]{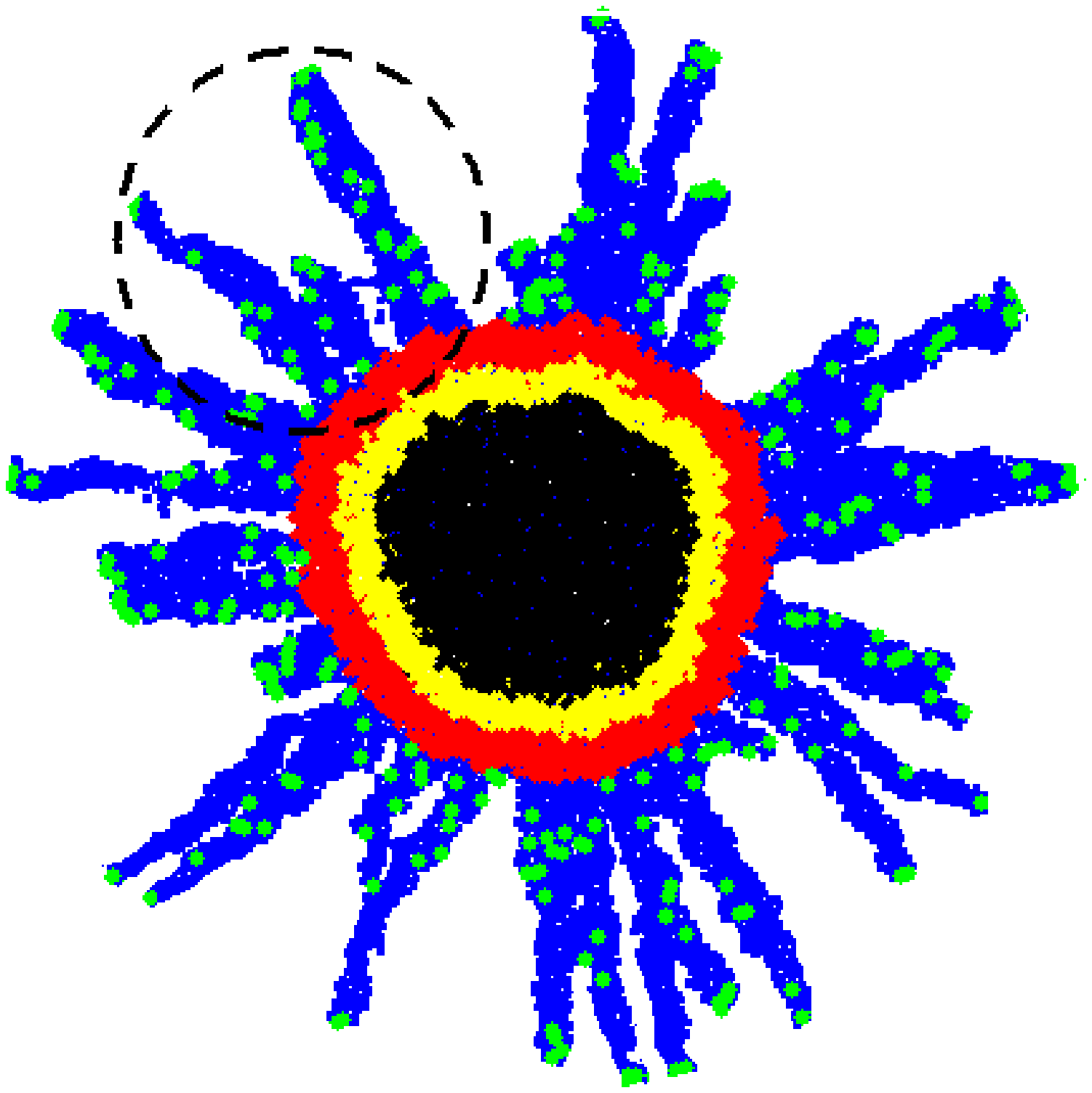} &
\includegraphics[height=4.25cm,keepaspectratio]{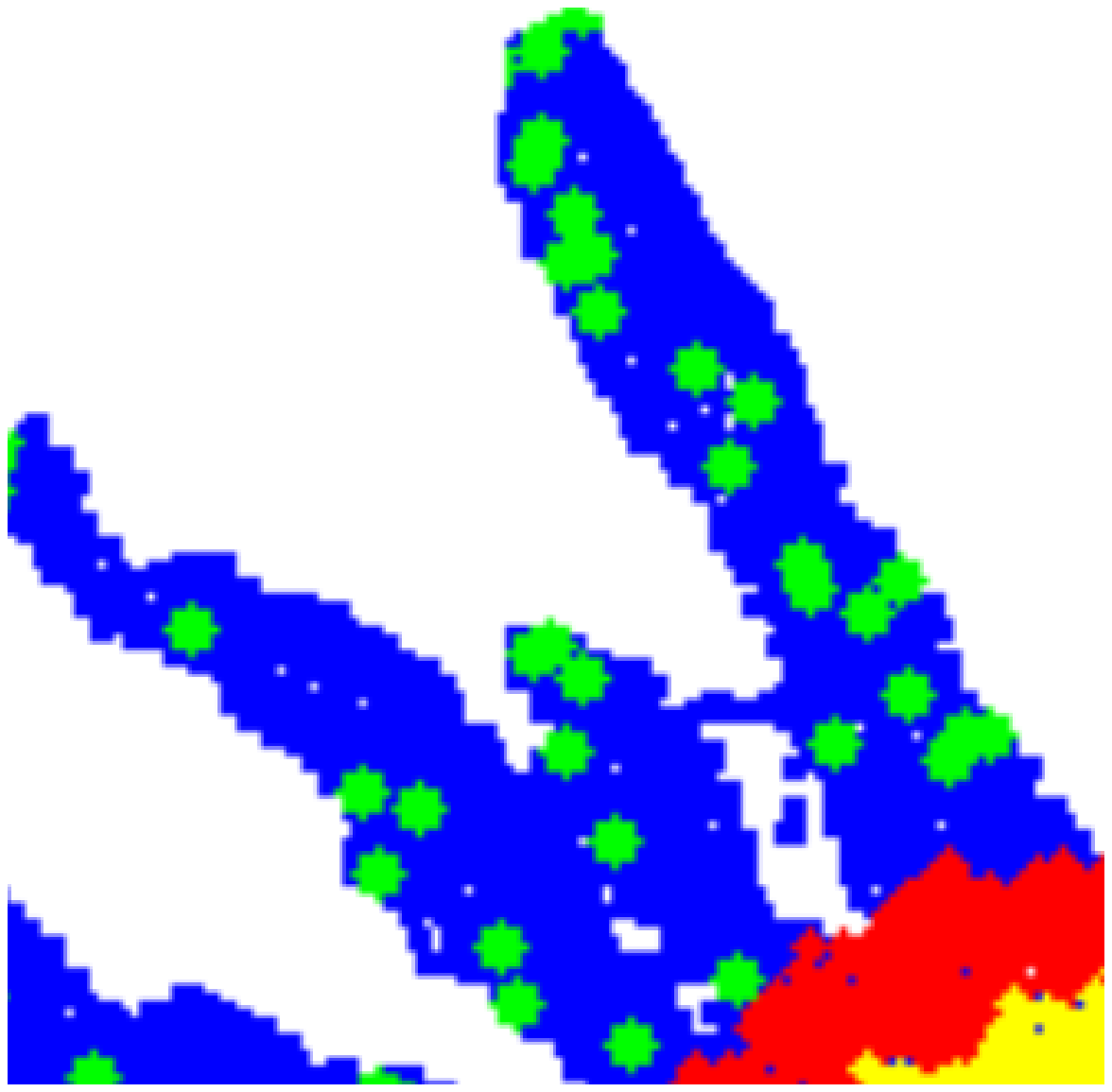} &
\includegraphics[height=4.25cm,keepaspectratio]{fig5c.eps} \\
\mbox{\bf (a)} & \mbox{\bf (b)} & \mbox{\bf (c)}
\end{array}$
\end{center}
\caption{{\bf Simulated invasive growth of MTS in vitro.}
(a) A snapshot of the simulated growing MTS at 24 hours after initialization.
The region circled is magnified in panel (b).
(b) A magnification of the circled region in panel (a). One can
clearly see that the invasive cells (green) are following each other to
form chains within the dendritic branches (blue), as observed in experiment \cite{deisboeck01}.
(c) Comparison of $\beta = A_I/A_T$ as a function of time associated
with the simulated MTS and the in vitro experimental data \cite{deisboeck01}.}
\label{fig_SimuMTS}
\end{figure}

Figure \ref{fig_SimuMTS}(a) and (b) respectively show the morphology
of simulated MTS and a magnification of its invasive branches with
increasing branch width towards the proliferative core. Specifically,
one can clearly see that within the branches, chains of cells are formed
as observed in experiments \cite{deisboeck01} (see Fig.~1). The invasive
cells tend to follow one another (which is termed ``homotype attraction'') since paths of degraded ECM are
formed by pioneering invasive cells and it is easier for other cells
to follow and enhance such paths than degrading ECM to create new paths by themselves.
In other words, invasive cells tend to take paths with ``least resistance''.
We note that no CA rules are imposed to force such
cellular behaviors. Instead, they are emergent properties that arise
from our simulations.

The ratio of invasion area over primary tumor area $\beta = A_{I}/A_{T}$
as a function of time for the simulated tumor is computed and compared to reported experimental
data \cite{deisboeck01} (see Fig.~\ref{fig_SimuMTS}(c)). One can clearly see that our simulation
results agree with experimental data very well. Moreover, the
deviation of $\beta(t)$ from a linear function of $t$ indicates
that the growth of primary tumor and the invasive branches are strongly coupled \cite{deisboeck01}
Other metrics for tumor morphology such as the
specific surface $s$ of the invasive pattern, the sphericity $\alpha$
of the primary tumor and the angular anisotropy metric $\psi$
for the invasive branches are computed from our simulation results
and from the image of invasive MTS in Fig.~1(a) at 24 hours after initialization.
The values are given in Table \ref{tab_MTS}, from which one can see again a good agreement.
Thus, we have shown that our CA model is both robust and quantitatively accurate with properly
selected parameters.


\subsection*{Simulated Invasive Growth in Heterogeneous Miroenvironments}



\begin{figure}[bthp]
\begin{center}
$\begin{array}{c@{\hspace{0.6cm}}c@{\hspace{0.6cm}}c}\\
\includegraphics[height=4.25cm,keepaspectratio]{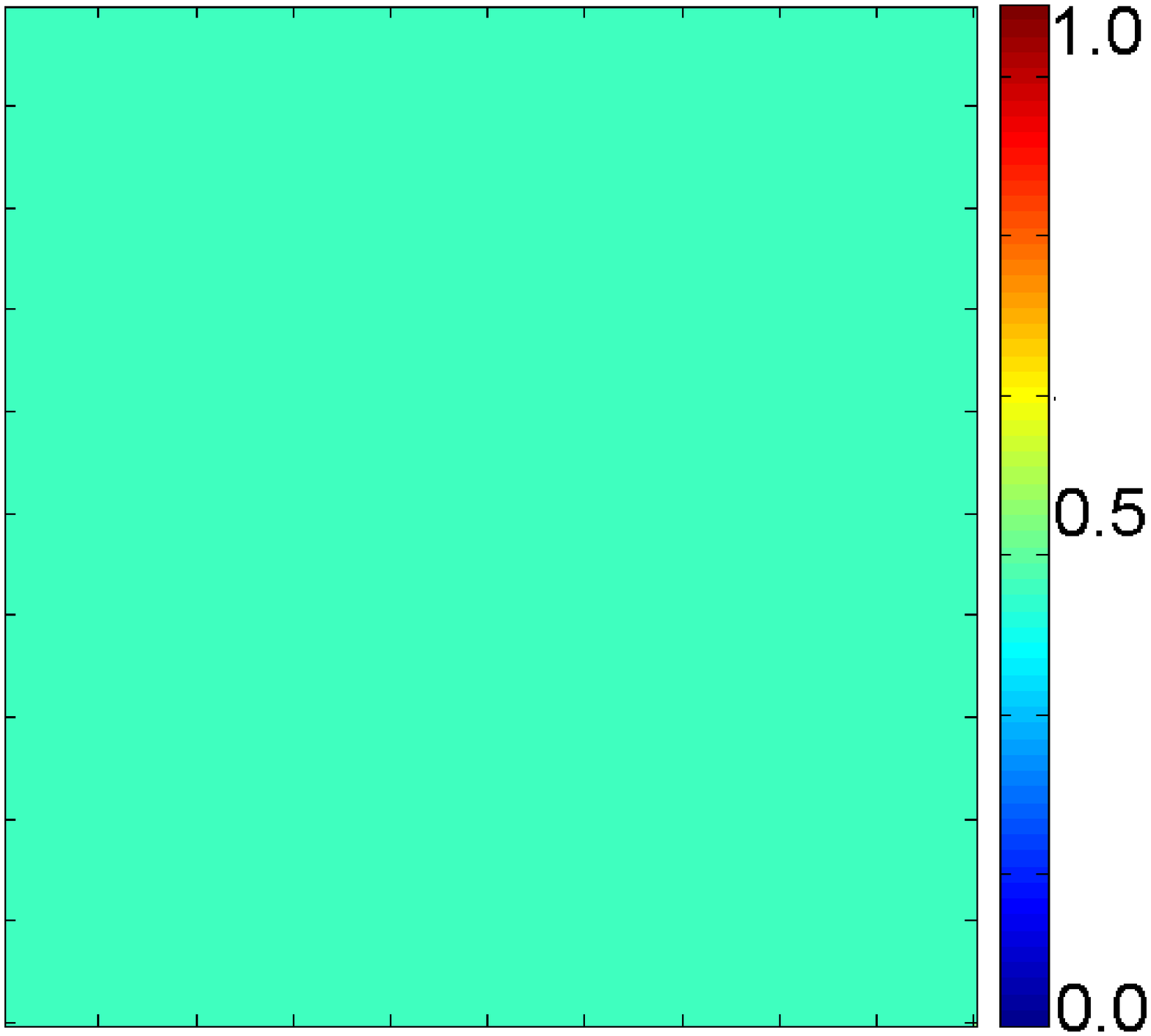} &
\includegraphics[height=4.25cm,keepaspectratio]{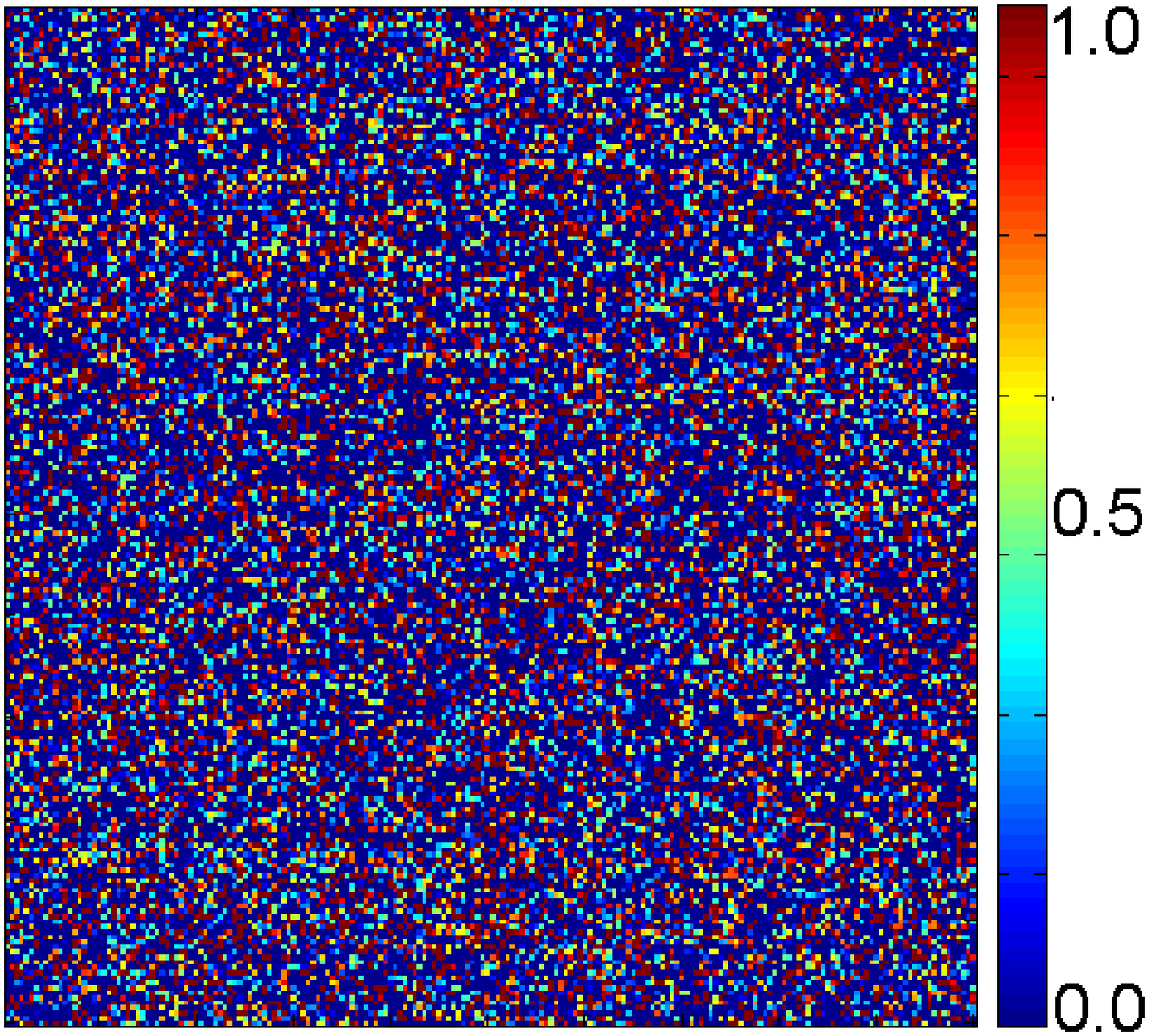} &
\includegraphics[height=4.25cm,keepaspectratio]{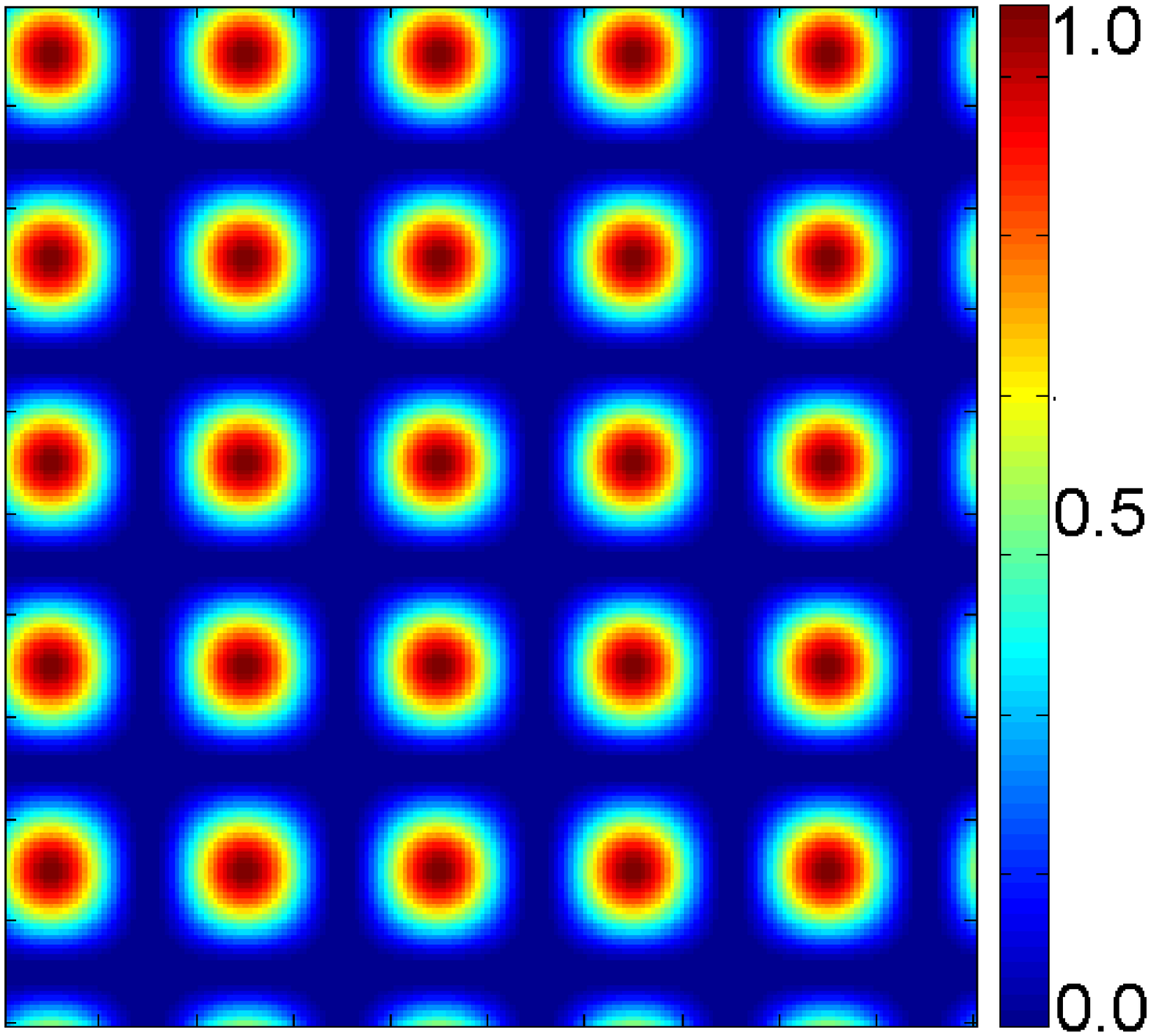} \\
\mbox{\bf (a)} & \mbox{\bf (b)} & \mbox{\bf (c)}
\end{array}$
\end{center}
\caption{{\bf Different distributions of ECM densities.}
(a) Uniform distribution. (b) Random distribution, i.e.,
the value of $\rho_{\mbox{\tiny{ECM}}}$ is completely independent of
$\rho_{\mbox{\tiny{ECM}}}$ values of other automaton cells.
(c) Sine-like distribution defined by Eq.~(\ref{eq_sine})
to mimic the obstacles for a growing tumor.}
\label{fig_ECM}
\end{figure}

Having verified the robustness and predictive capacity of our CA
model, we now consider three types of ECM density distributions,
i.e., homogeneous, random and sine-like, to systematically study
the effects of microenvironment heterogeneity on invasive tumor
growth (see Fig.~\ref{fig_ECM}). These ECM density distributions
represent real host microenvironments in which a tumor grows.
(Details about these ECM distributions are given in the following
sections.) Again, the boundary of the growth-permitting region is
considered to be vascularized with a constant radially symmetric
nutrient/oxygen gradient in the growth-permitting region pointing
to the tumor center. We note that although generally the
nutrient/oxygen concentration field in vivo is more complicated,
previously numerical studies that considered the exact evolution
of nutrient/oxygen concentrations have shown a decay of the
concentrations toward the tumor center \cite{anderson05, anderson06}. Since the directions of
cell motions are determined only by the nutrient/oxygen gradient, our
constant-gradient approximation is a very reasonable one.

In the beginning, a proliferative tumor cell is introduced at the
center of the growth-permitting region and tumor growth is
initiated. The growth parameters for the primary tumors in all
cases studied here are the same and are given in Table
\ref{tab_Param}. The invasiveness parameters and ECM densities are
variable and specified in each case separately. The values of the
growth parameters  for the CA model were chosen to be  consistent
with GBM  data from the medical literature \cite{kansal00a,
jana08}. Specifically, the value of the base probability of
division is $p_0 = 0.192$, which corresponds to a cell doubling
time of 4 days \cite{hoshino79, pertuiset85}. This value is used
for all of the cases of invasive growth that follow. Since our CA
model takes into account general microscopic tumor-host
interactions, we expect that the general growth dynamics and
emergent behaviors predicted by the model will qualitatively apply
to other solid tumors. We note that all of the reported growth dynamics and emergent
properties of the simulated tumors for any specific set of growth
and invasiveness parameters are repeatedly observed in 25
independent simulations.

\subsubsection*{Effects of Cellular Motility}
We first simulate the growth of malignant tumors with different degrees of
invasiveness in a homogeneous ECM with $\rho _{\mbox{\tiny{ECM}}} = 0.45$. In particular,
we consider three invasive cases with the same mutation rate $\gamma = 0.05$
and ECM degradation ability $\chi = 0.9$, but different cell motility
$\mu = 1, 2, 3$. A non-invasive growth case (i.e., $\gamma = 0$)
in the same microenvironment ($\rho_{\mbox{\tiny{ECM}}} = 0.45$) is also studied for comparison purposes.

\begin{figure}[bthp]
\begin{center}
$\begin{array}{c@{\hspace{0.6cm}}c@{\hspace{0.6cm}}c@{\hspace{0.6cm}}c}\\
\includegraphics[height=3.25cm,keepaspectratio]{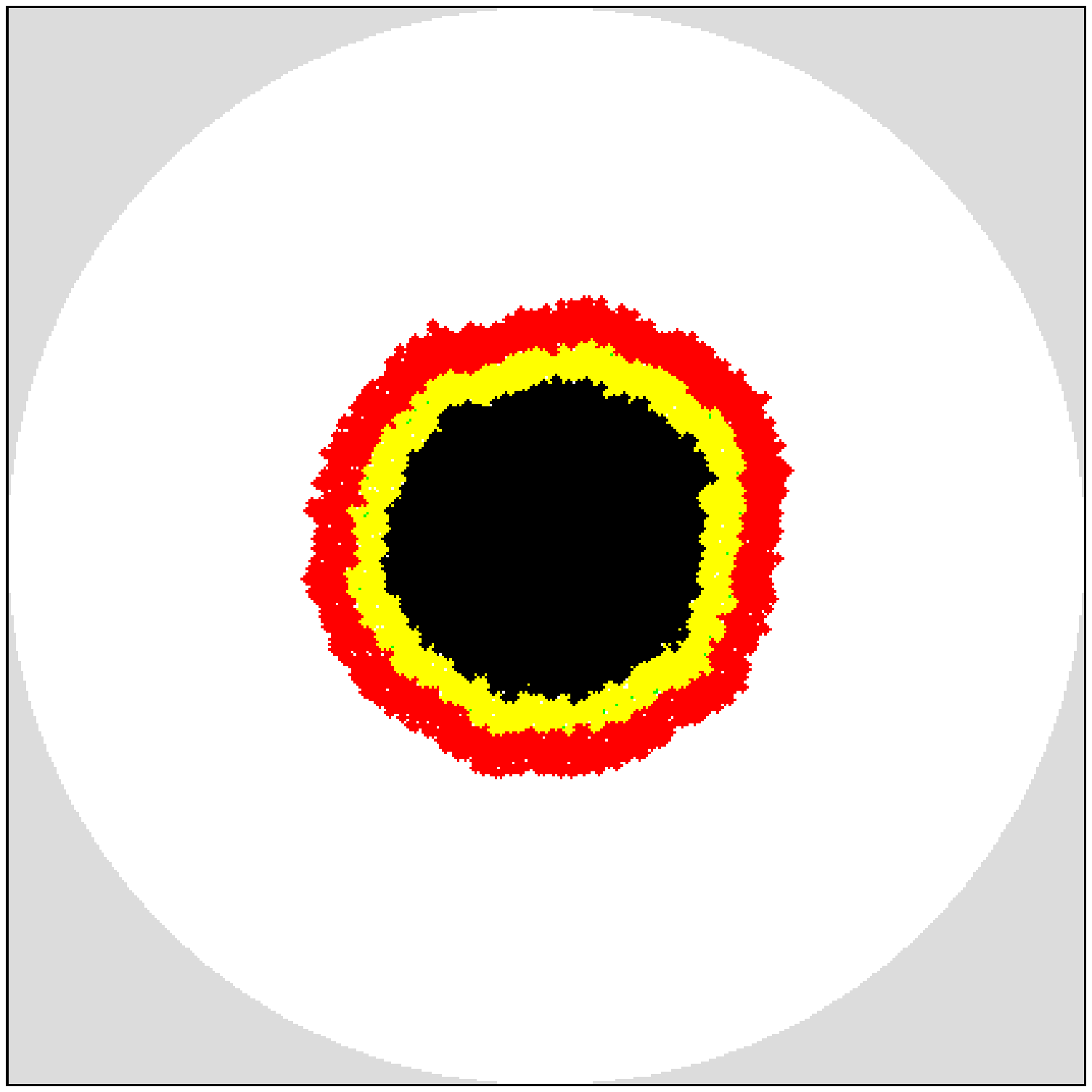} &
\includegraphics[height=3.25cm,keepaspectratio]{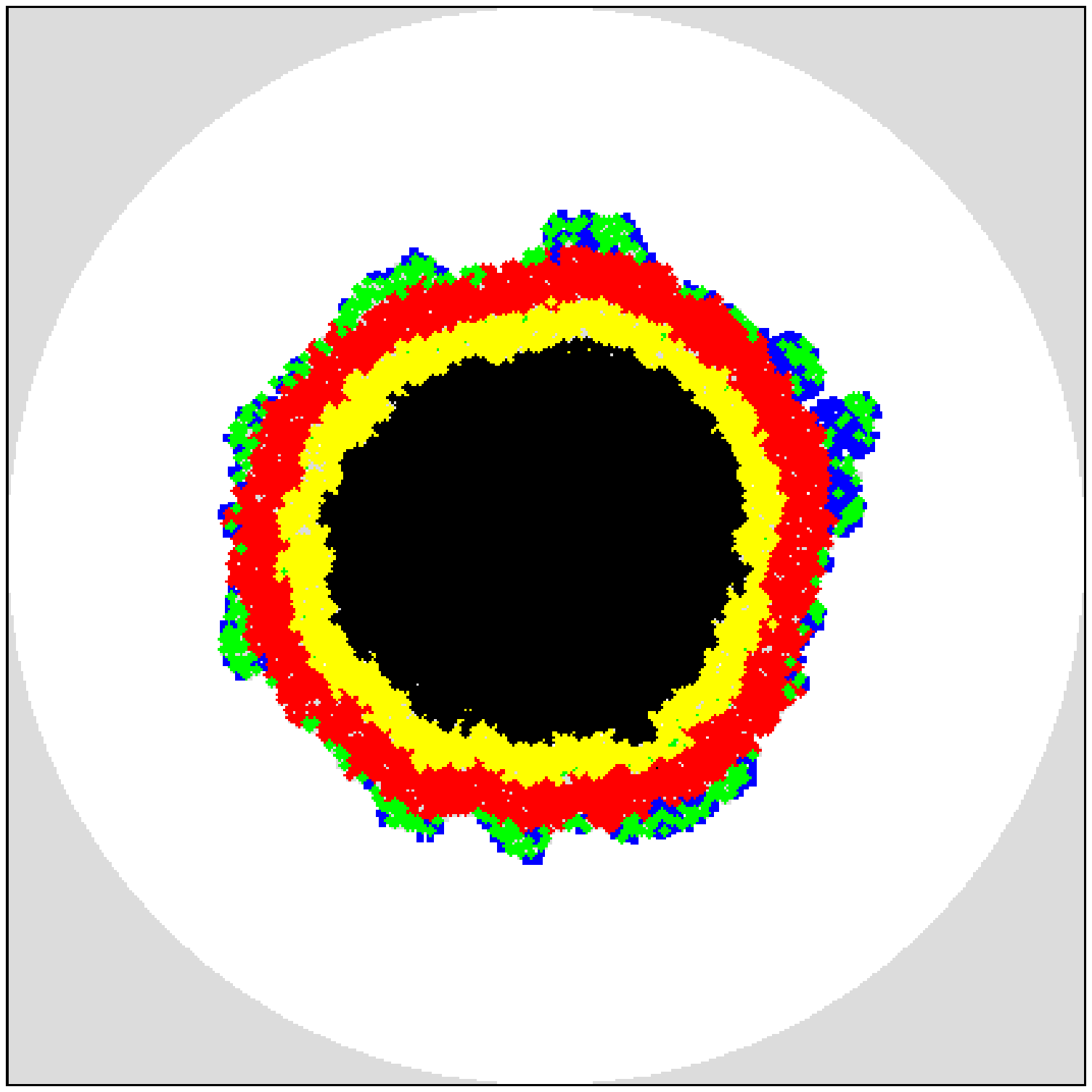} &
\includegraphics[height=3.25cm,keepaspectratio]{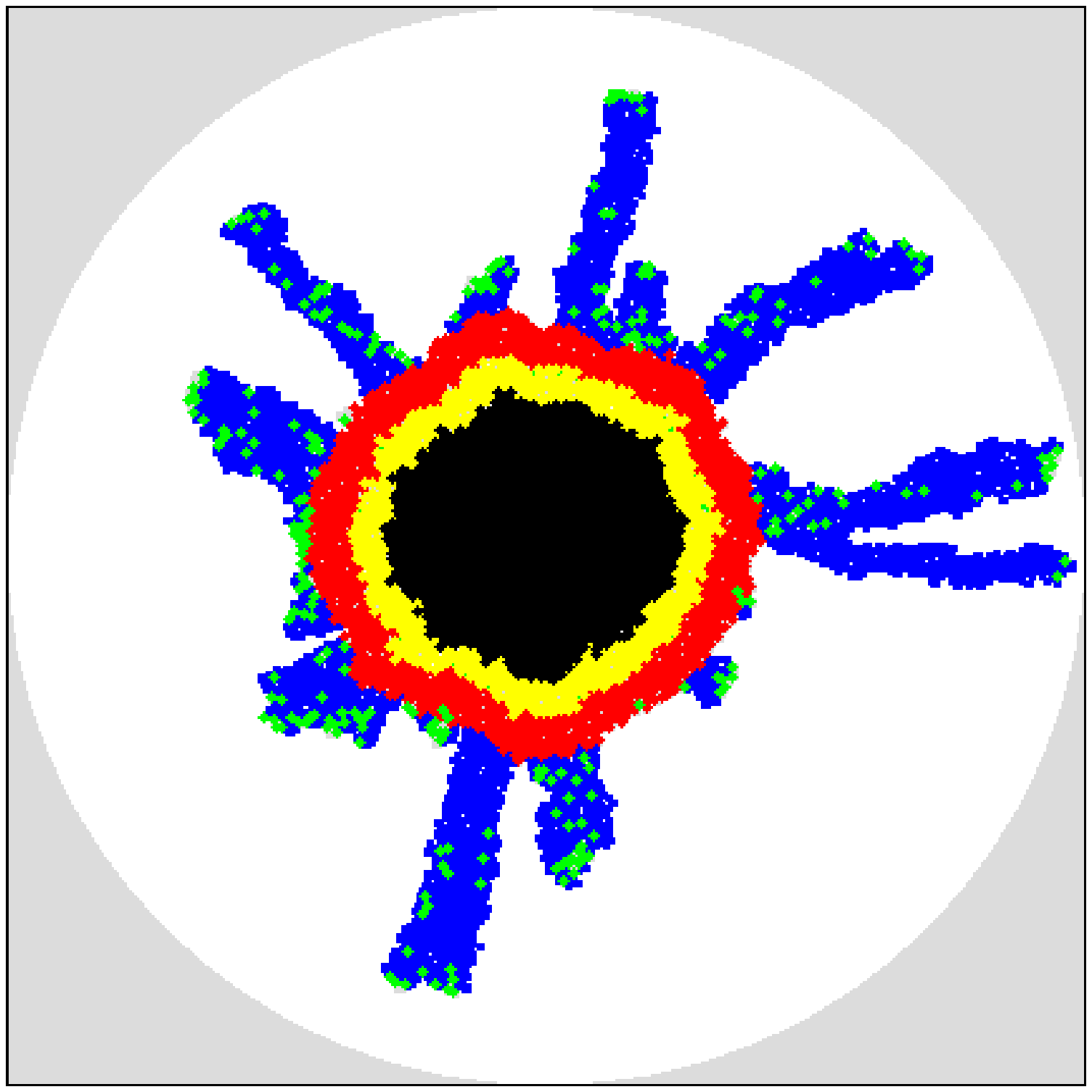} &
\includegraphics[height=3.25cm,keepaspectratio]{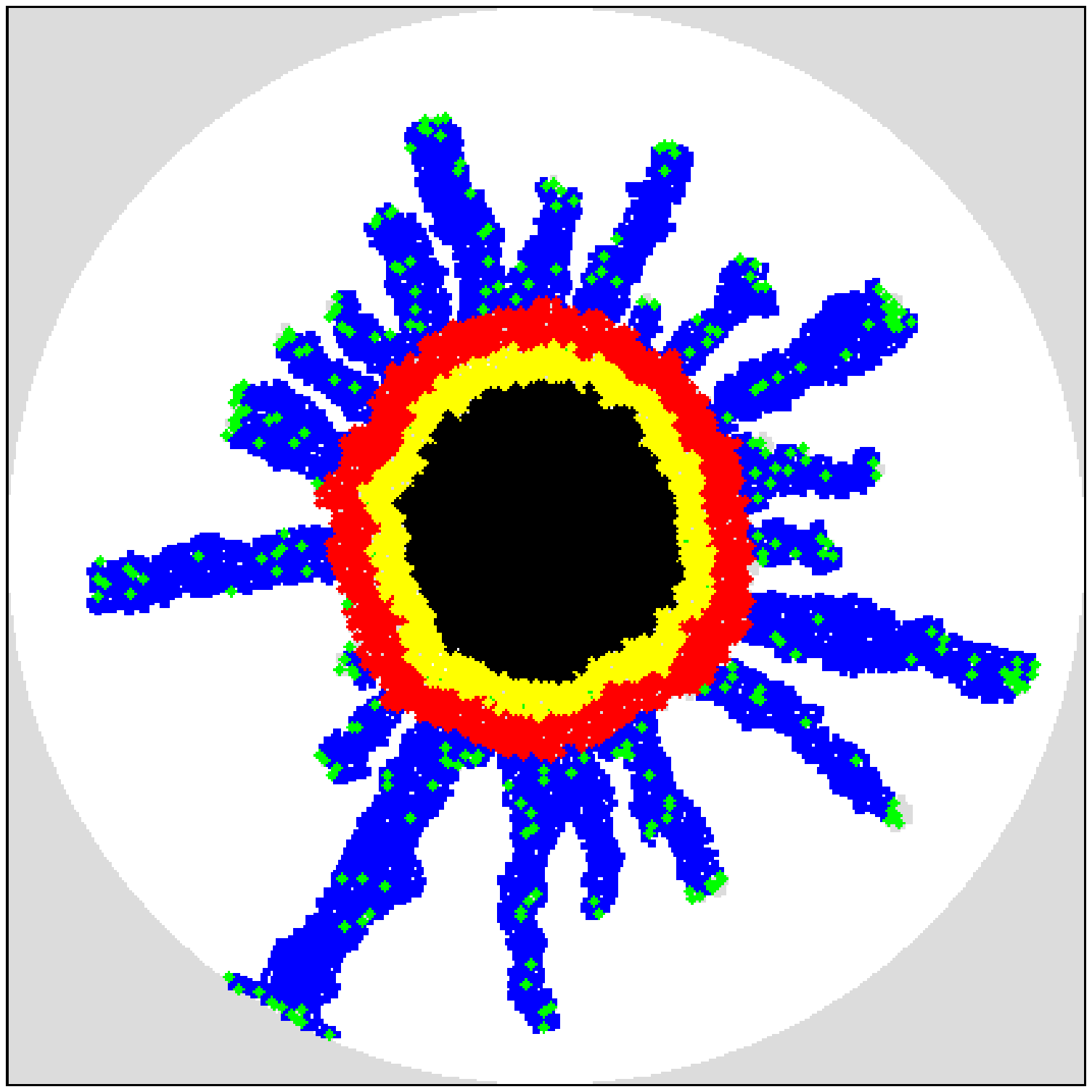} \\
\mbox{\bf (a)}~\mu = 0 & \mbox{\bf (b)}~\mu = 1& \mbox{\bf (c)}~\mu=2 & \mbox{\bf (d)}~\mu =3
\end{array}$
\end{center}
\caption{{\bf Simulated growing tumors in homogeneous ECM with $\rho_{\mbox{\tiny{ECM}}} = 0.45$
on day 100 after initiation.} For the invasive growth, the mutation rate is $\gamma = 0.05$
and ECM degradation ability is $\chi = 0.9$,
(a) Tumor cells are noninvasive, i.e., $\gamma = 0$.
(b) Invasive tumor with cellular motility $\mu=1$.
(c) Invasive tumor with cellular motility $\mu=2$.
(d) Invasive tumor with cellular motility $\mu=3$.
Note that the size of the primary tumor whose growth is facilitated by the
concentric-like shell formed by clumpped invasive cells (b) is much larger
than the other cases. Invasive cells with a larger motility
lead to more dendritic invasive branches. }
\label{fig_homo}
\end{figure}

Figure~\ref{fig_homo} shows the simulated growing tumors 100 days
after initiation (plots showing the full growth history of the
tumors are given in the Supplementary Information). The computed
metrics for tumor morphology are given in Table \ref{tab_homo}.
The primary tumors for both invasive [Figs.~\ref{fig_homo}(b),(c)
and (d)] and non-invasive [Fig.~\ref{fig_homo}(a)] cases develop
necrotic and quiescent regions. For invasive tumors, when the cell
motility is small (i.e., $\mu=1$), the invasive cells do not form
dendritic invasive branches but rather clump near the outer border
of the proliferative rim [see Fig.~\ref{fig_homo}(b)], forming
bumpy invasive concentric-like shells with relatively small
specific surface (e.g., $s = 1.09$ on day 100). Such invasive
shells significantly enhance the growth the primary tumor, i.e.,
the size of the primary tumor in Fig.~\ref{fig_homo}(b) is much
larger than in Figs.~\ref{fig_homo}(a), (c) and (d). A
quantitative comparison of the tumor sizes is given in the
Supplementary Information.

By contrast, for larger cell motility, long dendritic invasive
branches are developed as manifested by the large specific surface
(e.g., $s = 7.89$ on day 100 and $s = 9.73$ on day 120 for
$\mu=3$). In particular, one can clearly see that within the
branches the cells tend to follow one another to form chains, as
observed in experiments \cite{deisboeck01}. We emphasize that no
rules are imposed to force the cells to follow on another in our
CA model. This homotype attraction is purely due to the mechanical
interaction between the invasive cells and the ECM, i.e., once a
path of invasion is established by a leading invasive cell (by
degrading the ECM), other invasive cells nearby turn to follow and
enhance this path since the resistance for migration is minimized
on a existing path. Furthermore, we can see that larger cell
motility (i.e., high malignancy) leads to more invasive branches
[see Figs.~\ref{fig_homo}(c) and (d)] and thus, a larger specific
area of the invasive pattern.


\subsubsection*{Effects of the ECM Rigidity}



It is not very surprising that isotropic tumor shapes and
and invasive patterns are developed in a homogeneous ECM with relative low density
(i.e.,the ECM is soft) compared
to the ECM degradation ability of the invasive tumor cells. However,
real tumors are rarely isotropic, primarily due to the host
microenvironment in which they grow, which we now explore.

\begin{figure}[bthp]
\begin{center}
$\begin{array}{c@{\hspace{0.6cm}}c@{\hspace{0.6cm}}c@{\hspace{0.6cm}}c}\\
\includegraphics[height=3.25cm,keepaspectratio]{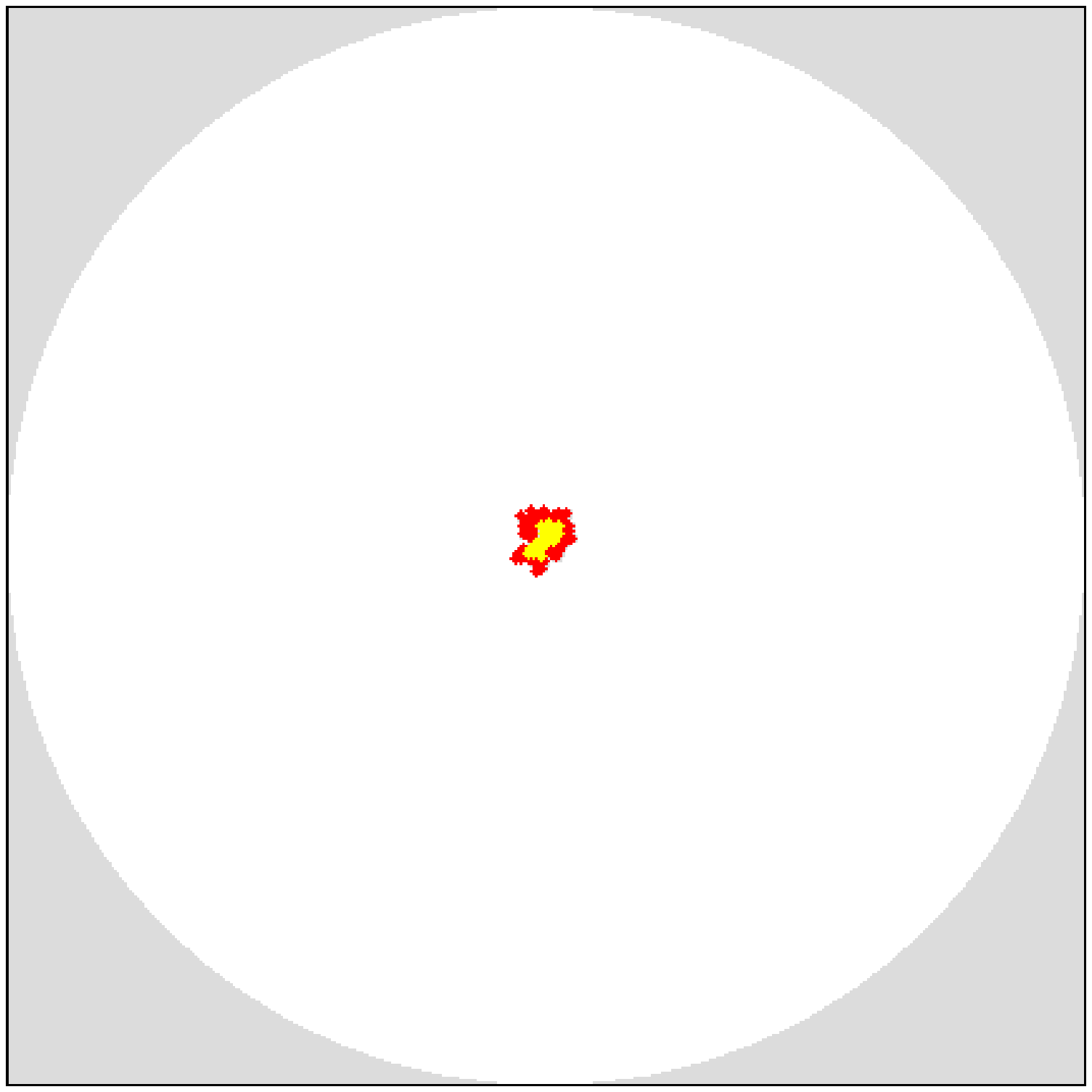} &
\includegraphics[height=3.25cm,keepaspectratio]{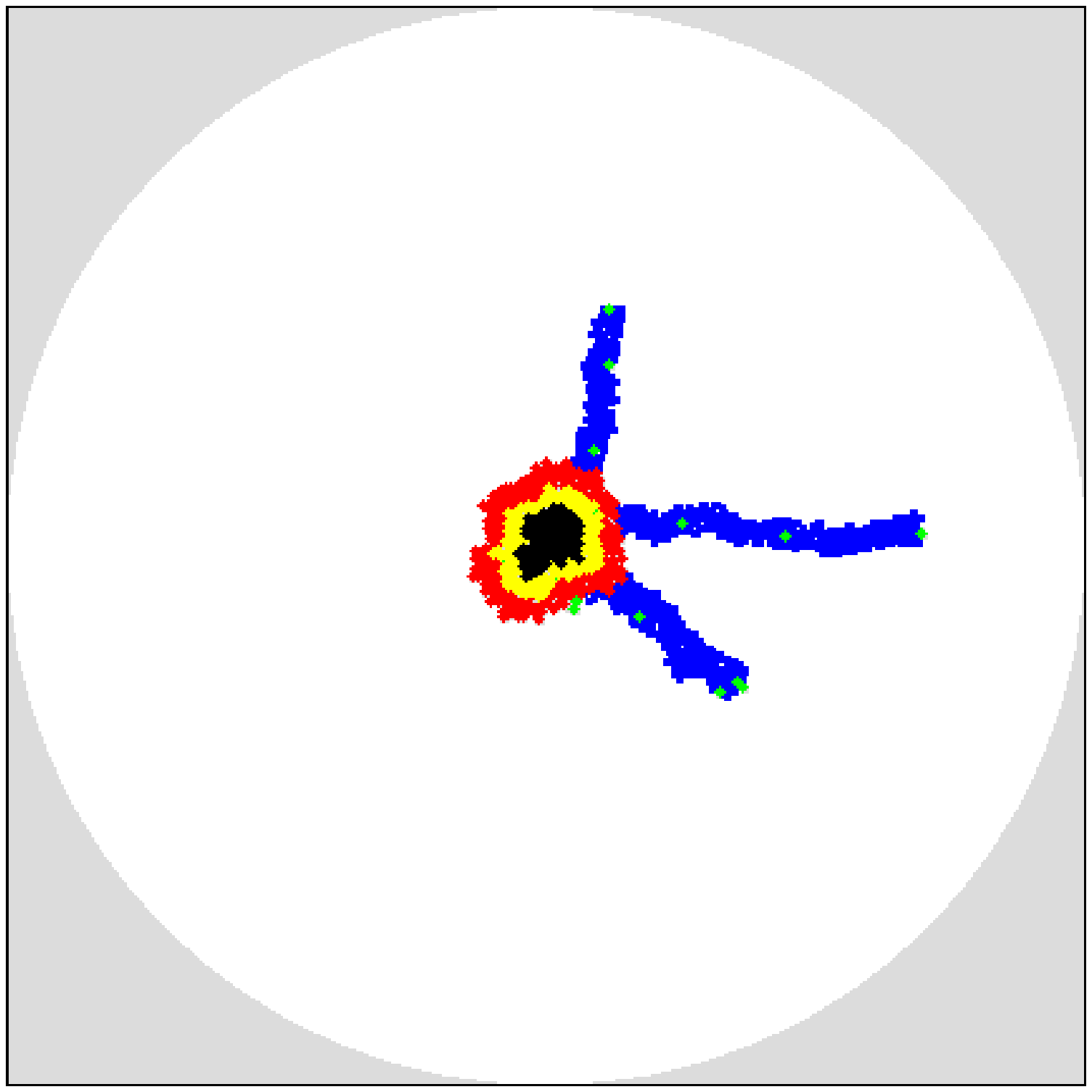} &
\includegraphics[height=3.25cm,keepaspectratio]{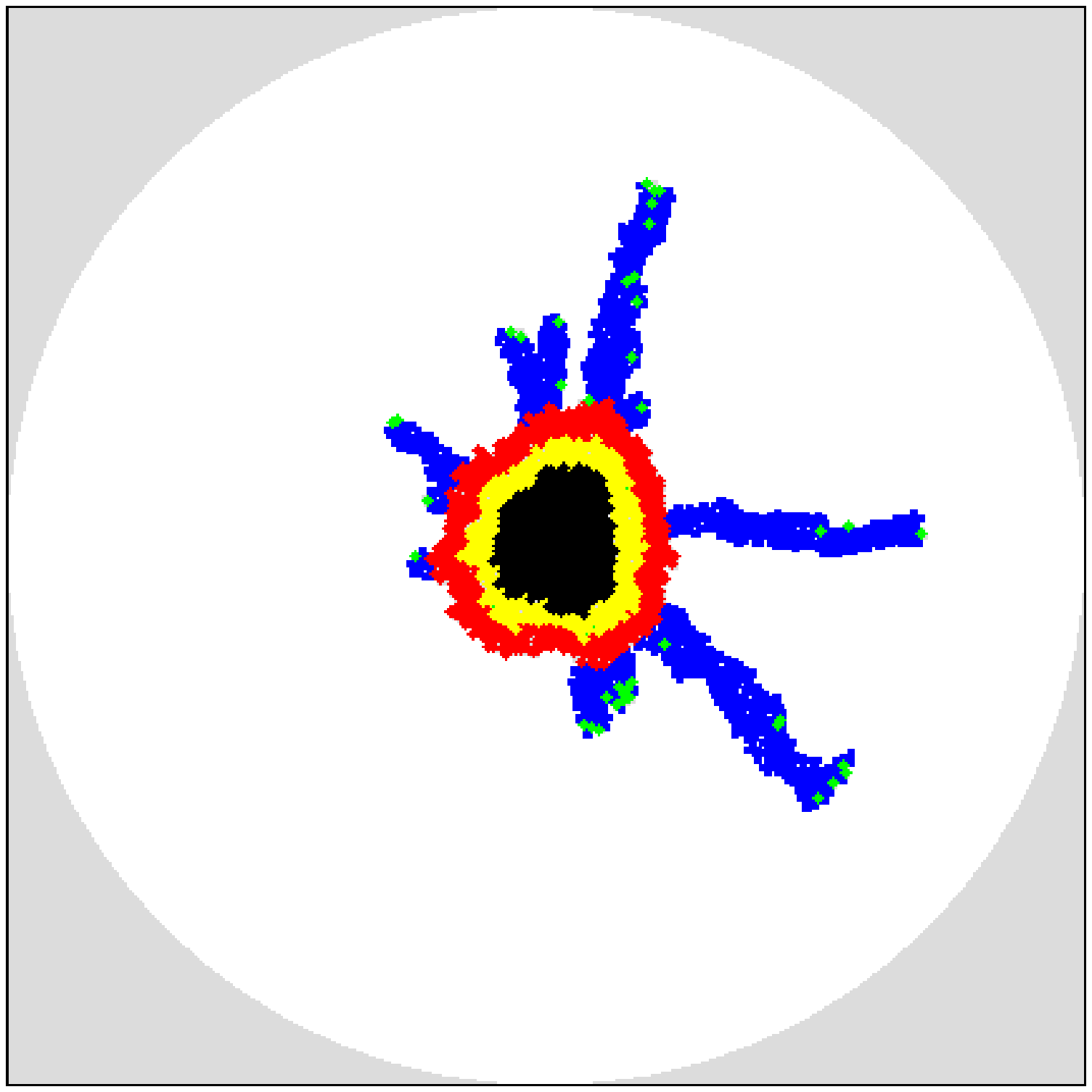} &
\includegraphics[height=3.25cm,keepaspectratio]{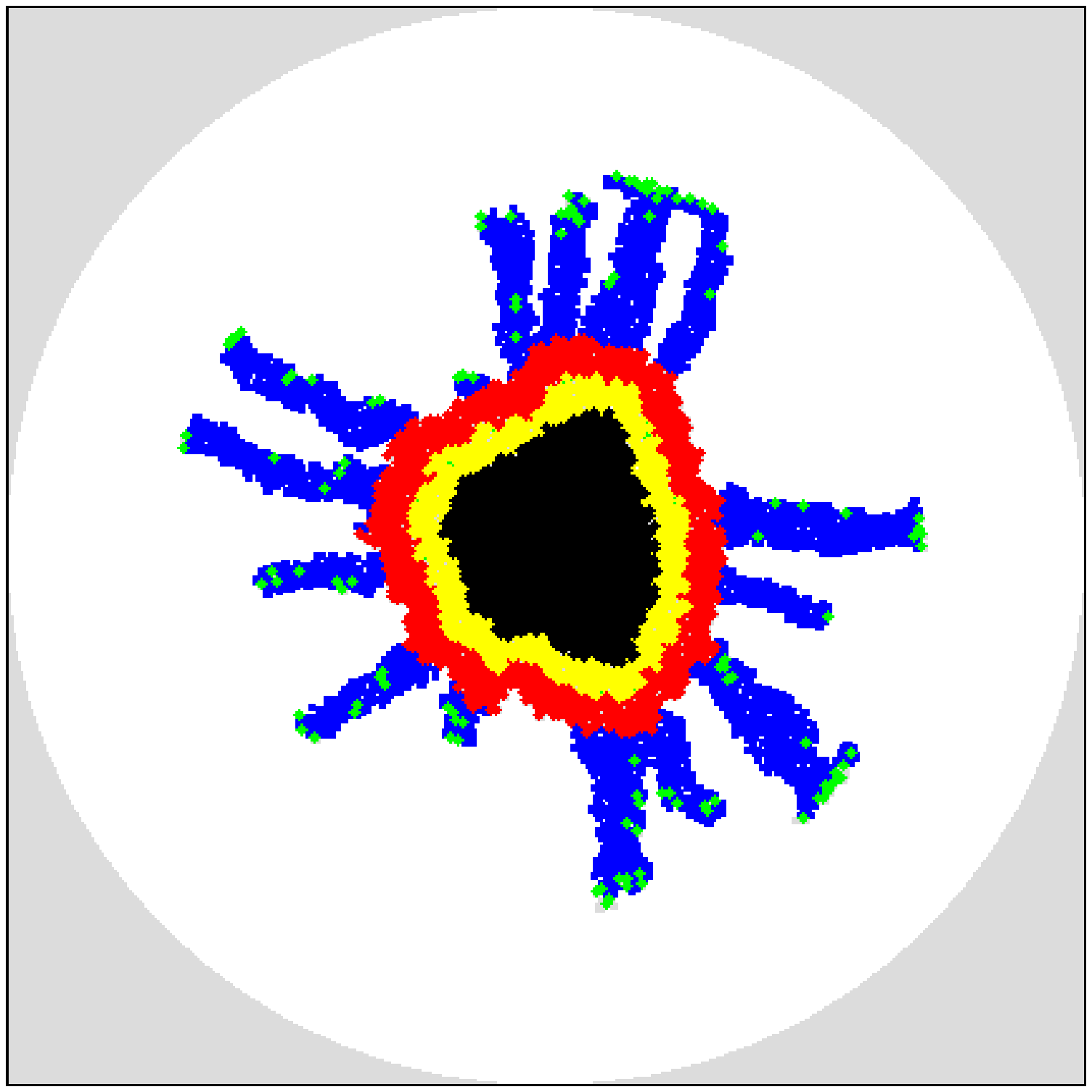} \\
\mbox{\bf (a)} & \mbox{\bf (b)} & \mbox{\bf (c)} & \mbox{\bf (d)}
\end{array}$
\end{center}
\caption{{\bf Evolution of simulated tumor in homogeneous ECM with $\rho_{\mbox{\tiny{ECM}}} = 0.85$}.
The mutation rate is $\gamma = 0.05$, the cell motility is $\mu = 3$
and ECM degradation ability is $\chi = 0.9$.
(a) Growing tumor on day 50. (b) Growing tumor on day 80.
(c) Growing tumor on day 100. (d) Growing tumor on day 120.
Note that although the ECM is homogeneous, due to its high rigidity, the primary
tumor develops an anisotropic shape with protrusions in the proliferation
rim caused by the invasive branches. Also note that the
invasive cells clump at the tips of certain invasive branches due
to the high ECM rigidity.}
\label{fig_rho}
\end{figure}

Consider the invasive growth of a tumor in a much more rigid ECM than
that in the previous section, i.e., $\rho_{\mbox{\tiny{ECM}}} =
0.85$. The invasiveness parameters used are $\gamma = 0.05$,
$\mu=3$ and $\chi = 0.9$. The snap shots of the growing tumor are
shown in Fig.~\ref{fig_rho} and the tumor morphology metrics are
given in Table \ref{tab_homo}. It can be clearly seen that both
the size of the primary tumor and the extent of its invasive
branches are much smaller than those of the tumors growing in a
softer ECM (see Fig.~\ref{fig_homo}). Importantly, although the
ECM is still homogeneous, due to its high rigidity, the primary
tumor develops an anisotropic shape with protrusions in the
proliferation rim caused by the invasive branches (e.g., $\alpha =
1.40$ and $\psi = 1.02$ on day 100). Since the invasive cells have
degraded the ECM either completely or partially along the invasive
branches, it is easier for the proliferative cells in the primary
tumor to take these ``weak spots'' than to push against the rigid
ECM themselves. Again, we emphasize that we do not force the cells
to behave this way by imposing special CA rules; this behavior
results purely from the mechanical interaction between the tumor
and its host and the coupling dynamics of invasive and
non-invasive tumor cells.

\subsubsection*{Effects of the ECM Heterogeneity: Random Distribution of ECM Density}

\begin{figure}[bthp]
\begin{center}
$\begin{array}{c@{\hspace{0.6cm}}c@{\hspace{0.6cm}}c@{\hspace{0.6cm}}c}\\
\includegraphics[height=3.25cm,keepaspectratio]{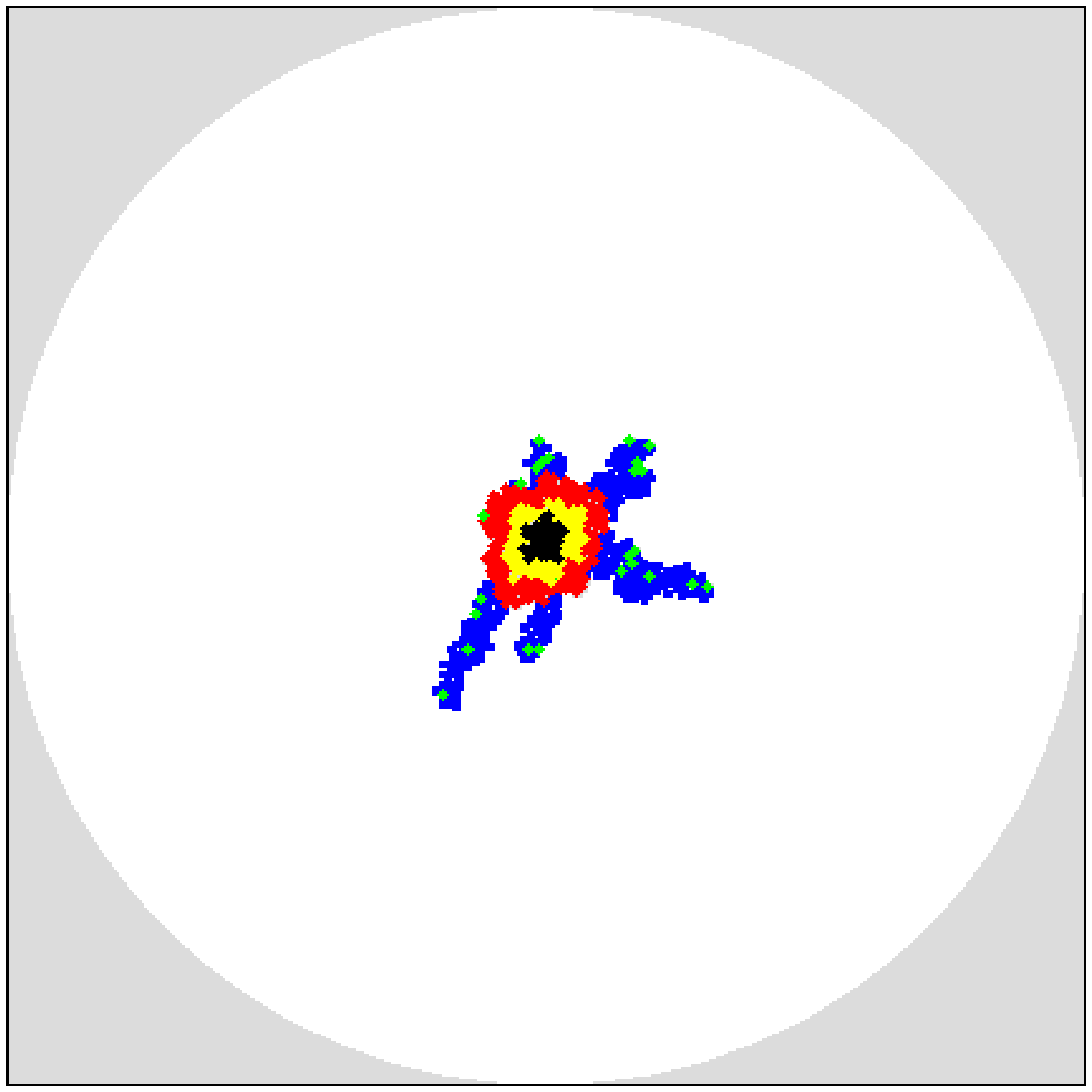} &
\includegraphics[height=3.25cm,keepaspectratio]{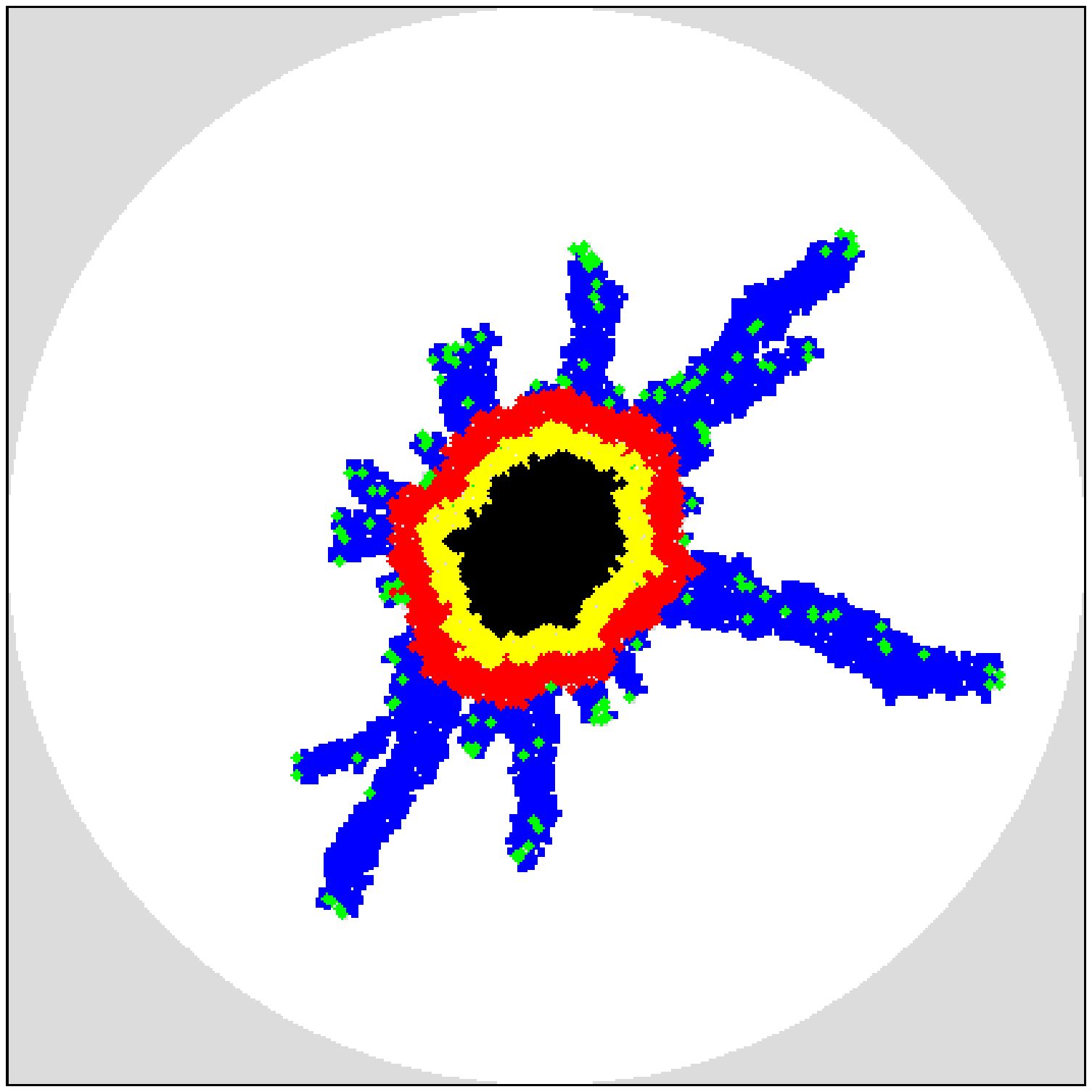} &
\includegraphics[height=3.25cm,keepaspectratio]{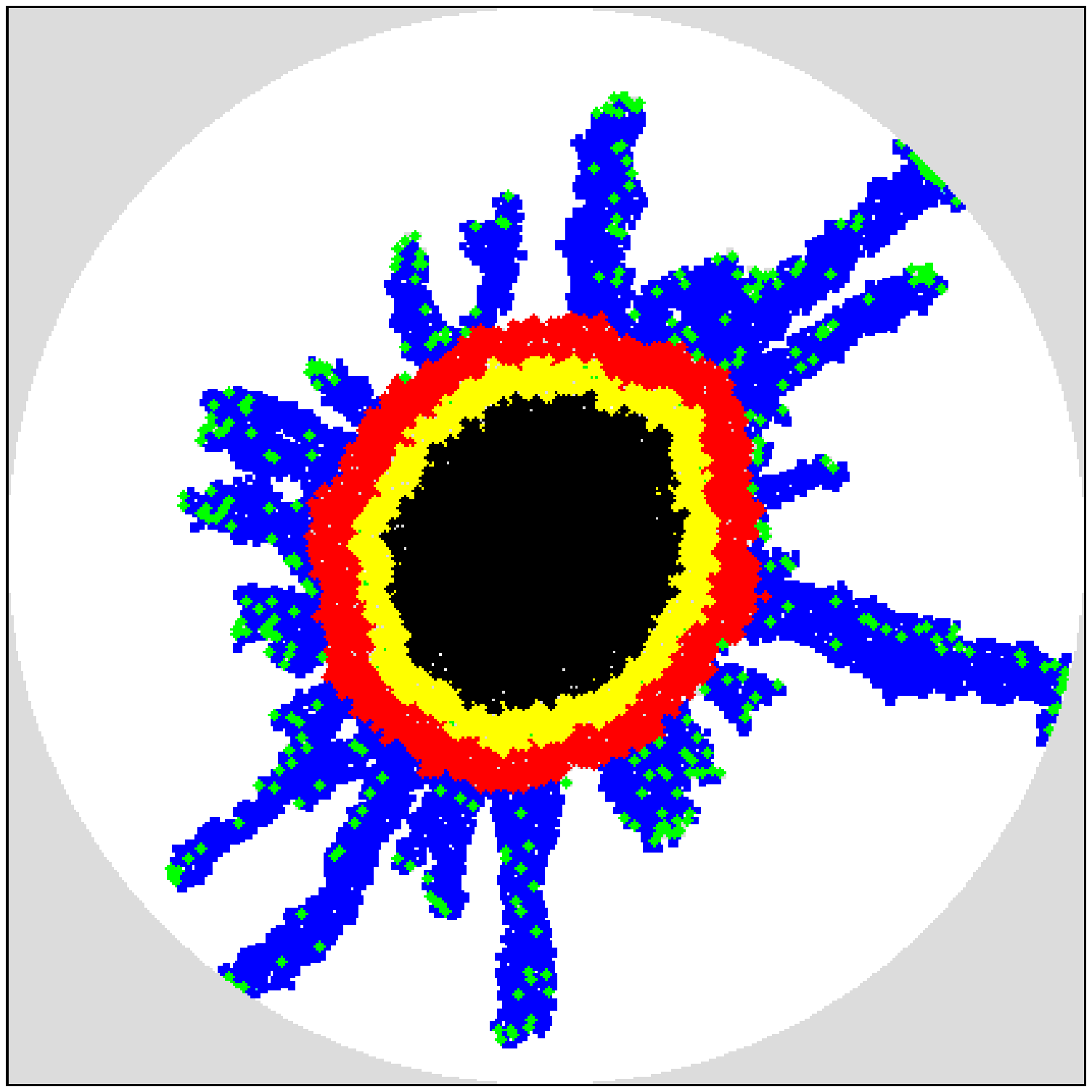} &
\includegraphics[height=3.25cm,keepaspectratio]{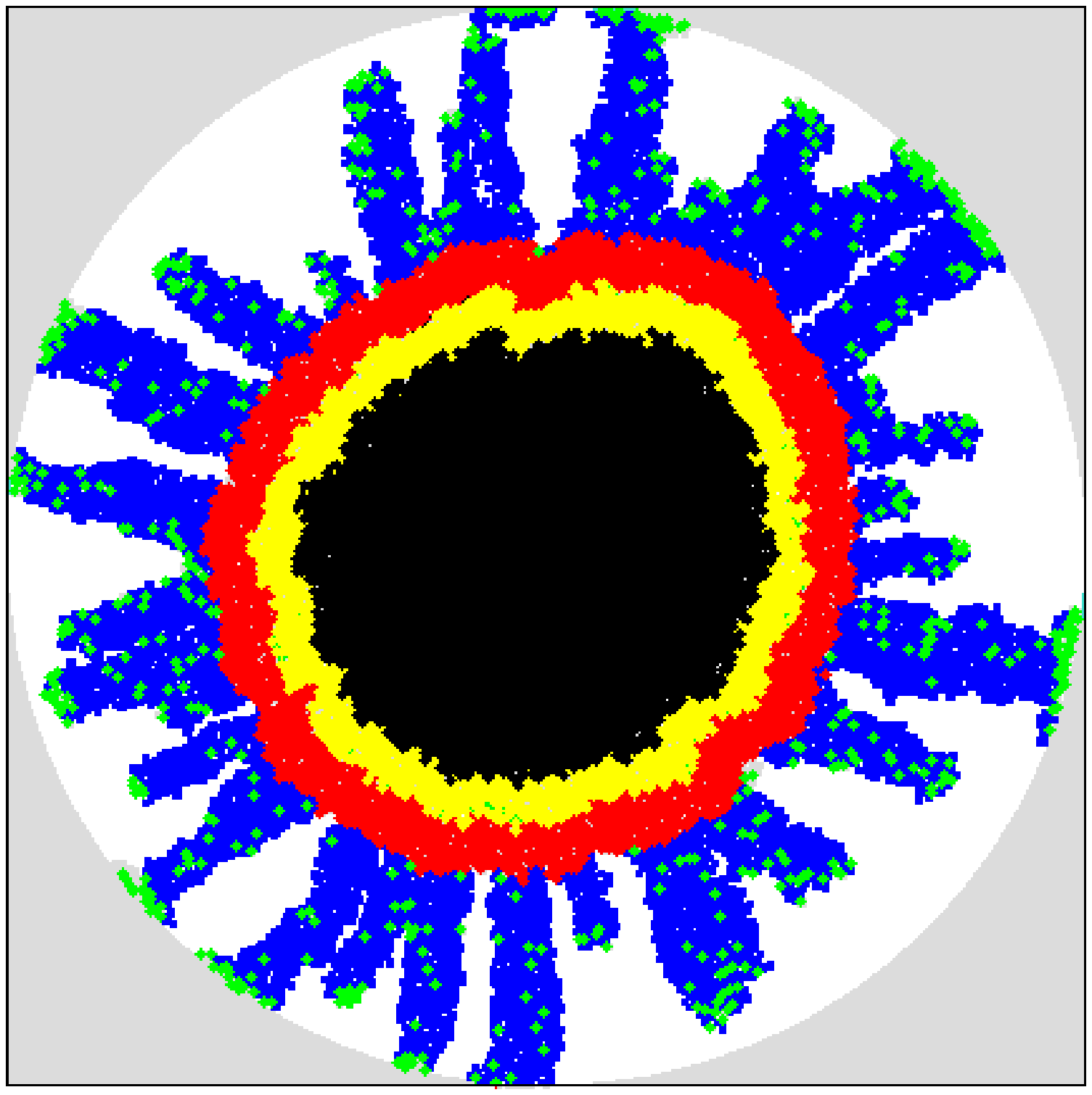} \\
\mbox{\bf (a)} & \mbox{\bf (b)} & \mbox{\bf (c)} & \mbox{\bf (d)}
\end{array}$
\end{center}
\caption{{\bf Evolution of simulated tumor in random ECM}.
The mutation rate is $\gamma = 0.05$, the cell motility is $\mu = 3$
and ECM degradation ability is $\chi = 0.9$.
(a) Growing tumor on day 50. (b) Growing tumor on day 80.
(c) Growing tumor on day 100. (d) Growing tumor on day 120.
Note that both the primary tumor and invasive pattern are
affected (i.e., becoming anisotropic) by the ECM heterogeniety
in the early growth stages. Also note that unlike the case
in Fig.6(d), the invasive cells clump at the tips of invasive
branches since they have reached the boundary of the growth-permitting region.}
\label{fig_rand}
\end{figure}

The real host microenvironment for tumors are far from homogeneous
in general. To investigate how heterogeneity of ECM affects the
tumor growth dynamics, we use a random distribution of ECM
density, i.e., for each ECM associated automaton cell, it density
$\rho_{\mbox{\tiny{ECM}}}$ is a random number uniformly chosen
from the interval $[0, ~1]$ (see Fig.~\ref{fig_ECM}(b)). The
invasiveness parameters used are $\gamma = 0.05$, $\mu=3$ and
$\chi = 0.9$ and snap shots of the growing tumor are shown in
Fig.~\ref{fig_rand}. The tumor morphology metrics are given in
Table \ref{tab_hetero}. Note that the primary tumor develops a
rough surface and slightly anisotropic shape in the early growth
stages (e.g.,  $\alpha = 1.32$ on day 50 and $\alpha = 1.34$ on
day 80), which reflects the ECM heterogeneity
[Figs.~\ref{fig_rand}(a) and (b)]. Since the characteristic
heterogeneity length scale is comparable to a single cell, its
effects are diminished (e.g.,  $\alpha = 1.18$ on day 100 and
$\alpha = 1.15$ on day 120) as the tumor grows larger and larger
[Fig.~\ref{fig_rand}(c) and (d)]. (In other words, on large length
scales, the ECM is still effectively homogeneous.) However, the
anisotropy in the invasive pattern (i.e., the extents of invasive
branches in different directions) still persists (e.g., $\psi =
0.64$ on day 100) even though the primary tumors almost resumes an
isotropic shape.



\subsubsection*{Effects of the ECM Heterogeneity: Sine-like Distribution of ECM Density}

To represent large-scale heterogeneities in the ECM, we use a sine-like
distribution of ECM density, i.e., for an automaton cell with centroid
$(x_1, x_2, \ldots, x_d)$, the associated ECM density is given by
\begin{equation}
\label{eq_sine}
\rho_{\mbox{\tiny{ECM}}}(x_1, \ldots, x_d) = \frac{1}{2^d}[\sin(\frac{2\pi x_1}{L})+1][\sin(\frac{2\pi x_2}{L})+1]\cdots[\sin(\frac{2\pi x_d}{L})+1],
\end{equation}
where $d$ is the spatial dimension and $L$ is the edge length of the
$d$-dimensional cubic simulation box. A two-dimensional sine-like
ECM distribution is shown in Fig.~\ref{fig_ECM}(c). The red spots correspond
to large $\rho_{\mbox{\tiny{ECM}}}$ and high ECM rigidity; they can be considered
as effective obstacles (e.g., brain ventricles) that hinder tumor growth.

\begin{figure}[bthp]
\begin{center}
$\begin{array}{c@{\hspace{0.6cm}}c@{\hspace{0.6cm}}c@{\hspace{0.6cm}}c}\\
\includegraphics[height=3.25cm,keepaspectratio]{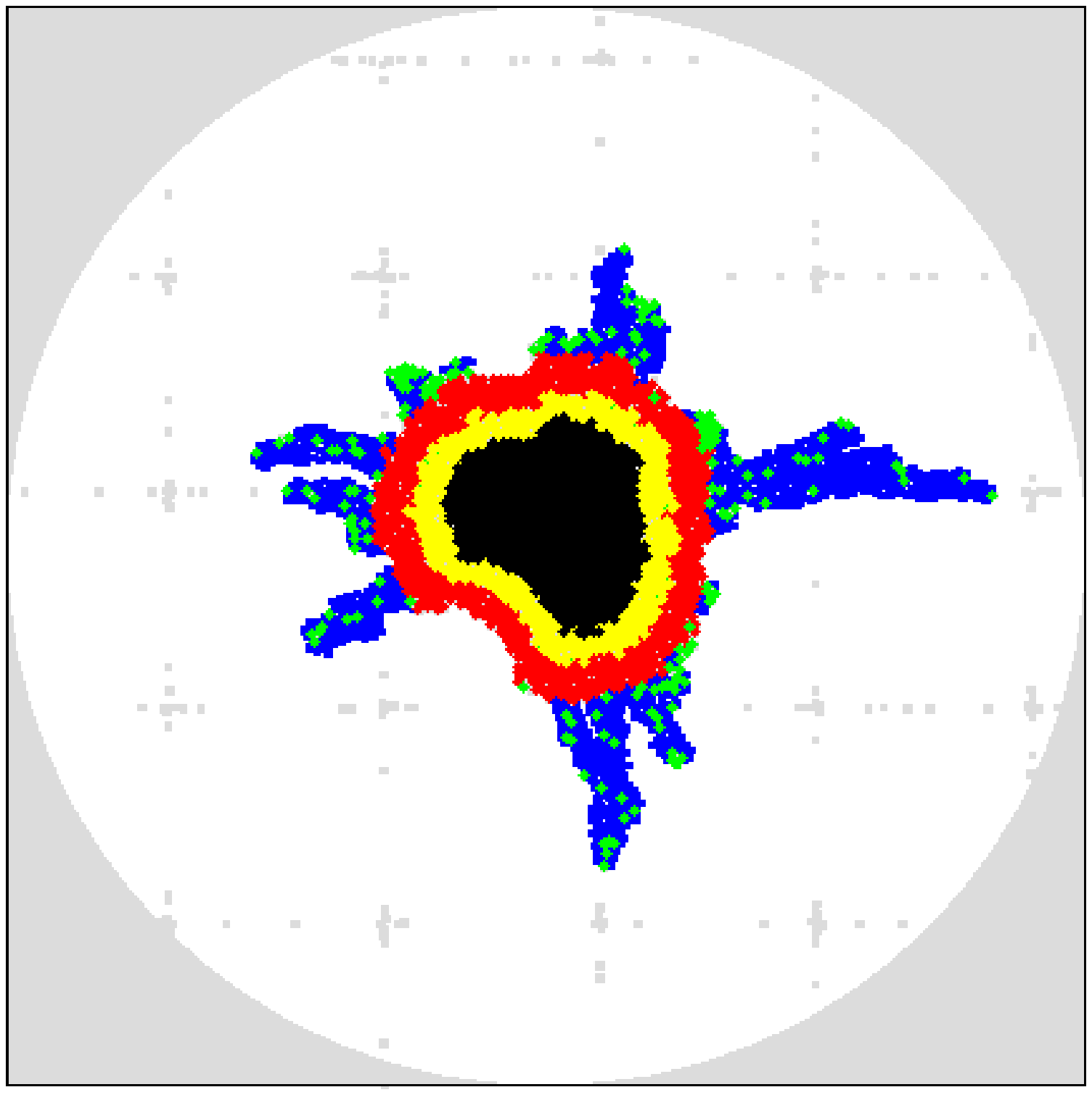} &
\includegraphics[height=3.25cm,keepaspectratio]{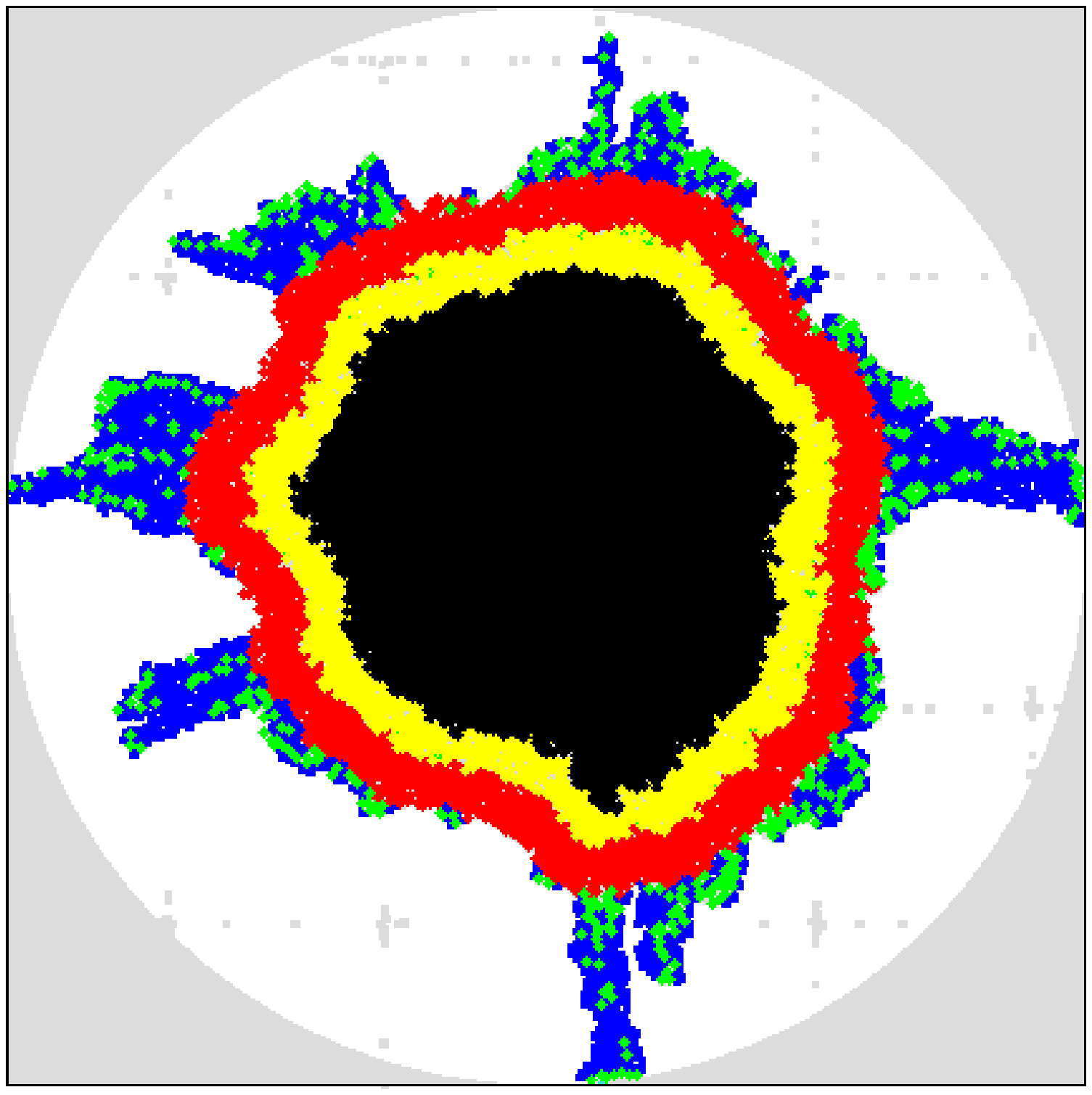} &
\includegraphics[height=3.25cm,keepaspectratio]{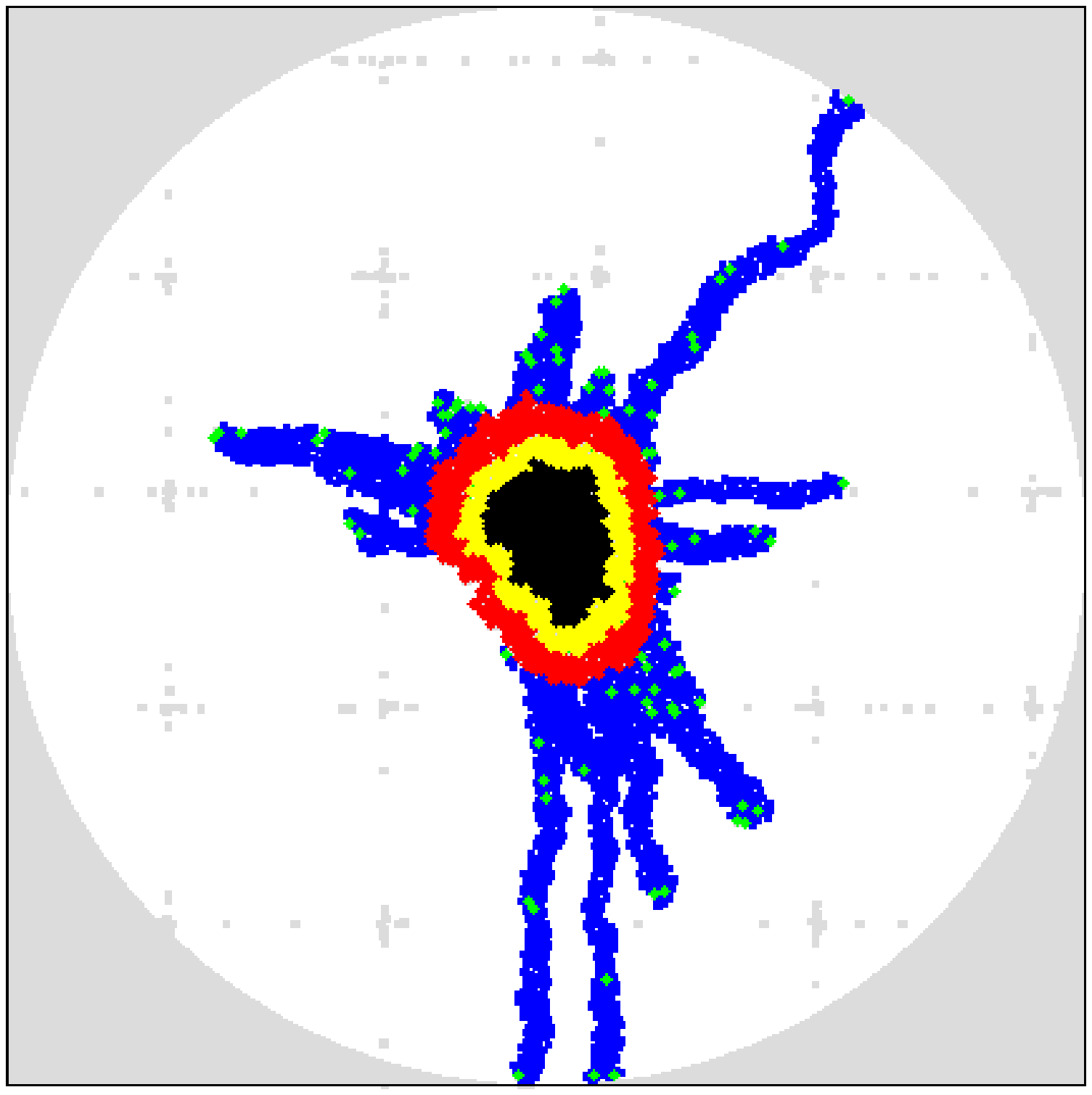} &
\includegraphics[height=3.25cm,keepaspectratio]{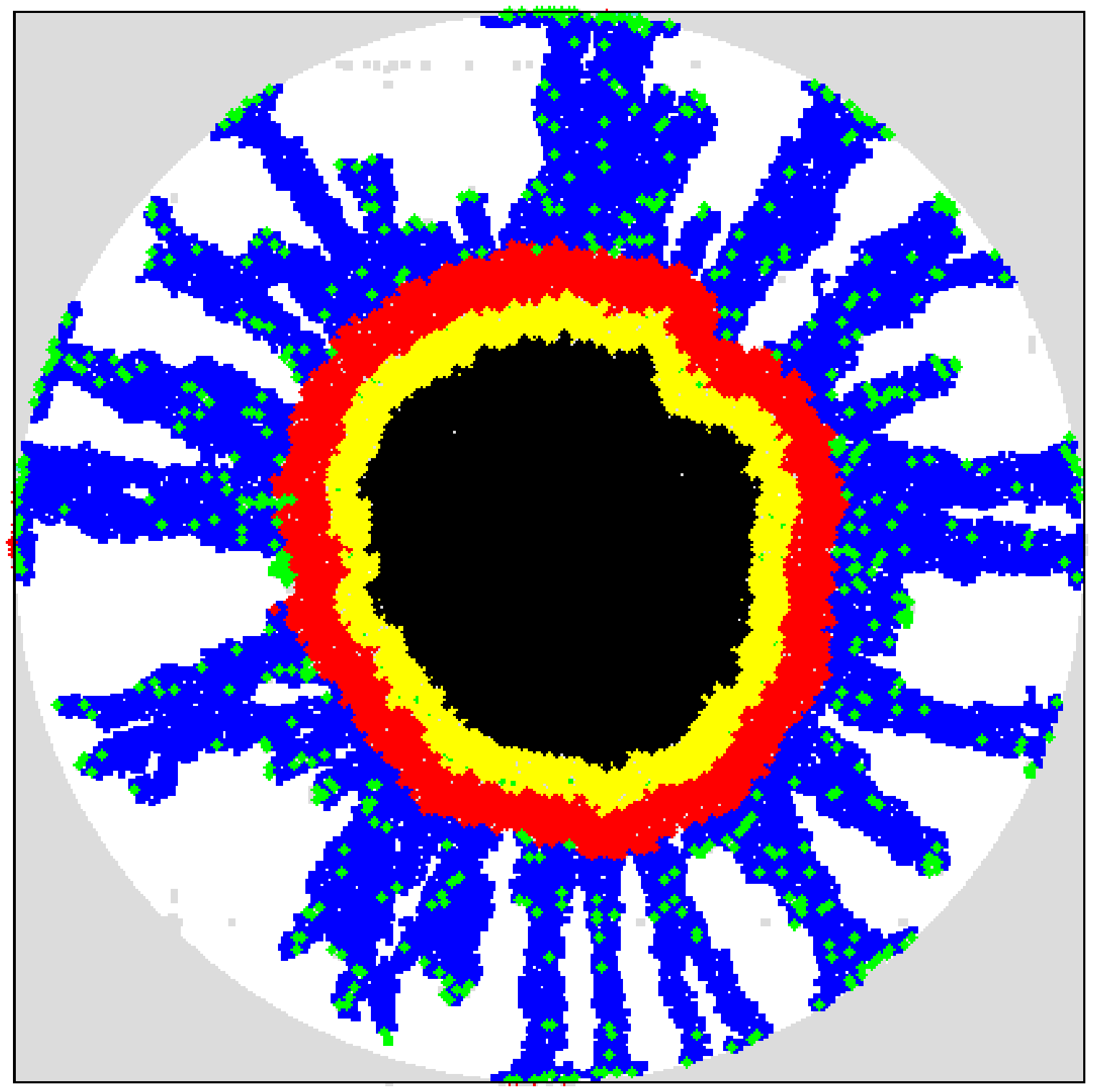} \\
\mbox{\bf (a)}~ \mu = 1& \mbox{\bf (b)}~\mu=1 & \mbox{\bf (c)}~\mu=3& \mbox{\bf (d)}~\mu=3
\end{array}$
\end{center}
\caption{{\bf Simulated growing tumors in sine-like ECM with different motilities ($\mu=1,3$)}.
The mutation rate is $\gamma = 0.05$ and ECM degradation ability is $\chi = 0.9$.
(a) Growing tumor with cell motility $\mu=1$ on day 80.
(b) Growing tumor with cell motility $\mu=1$ on day 120.
(c) Growing tumor with cell motility $\mu=3$ on day 80.
(d) Growing tumor with cell motility $\mu=3$ on day 120.
Note that both the primary tumor and invasive pattern
in the two cases are significantly affected by the ECM heterogeneity,
i.e., the tumors are highly anisotropic in shape and the
invasive branches clearly favor two orthogonal directions associated
with low ECM densities in the early growth stages.}
\label{fig_sine}
\end{figure}

Figures~\ref{fig_sine}(a),(b) and (c),(d) show the snap shots of
invasive tumors growing in the aforementioned ECM on day 80 and
day 120, with invasiveness parameters $\gamma = 0.05$, $\mu=1$ and
$\chi = 0.9$ and $\gamma = 0.05$, $\mu=3$ and $\chi = 0.9$,
respectively. The plots showing the full growth history is given
in Supplementary Information and the tumor morphology metrics are
given in Table \ref{tab_hetero}. We can see that in the early
growth stage, both the primary tumor and invasive pattern in the
two cases are significantly affected by the ECM heterogeneity. In
particular, the tumors are highly anisotropic in shape and the
invasive branches clearly favor two orthogonal directions
associated with low ECM densities (e.g., $\alpha = 1.61$, $\psi =
1.18$ for $\mu =1$ on day 80; and $\alpha = 1.67$, $\psi = 1.33$
for $\mu =3$ on day 80). For the case with large cellular
motility, anisotropy effects are diminished in later growth stages
($\alpha = 1.21$, $\psi = 0.23$ for $\mu =3$ on day 120). For
small cellular motility, anisotropy in both primary tumor shape
and the invasive pattern persists ($\alpha = 1.26$, $\psi = 0.98$
for $\mu =1$ on day 120). Furthermore, one can see that again
invasive cells with low motility significantly facilitate the
growth of the primary tumor. However, instead of forming ``bumpy''
concentric-like shells as in homogeneous ECM, the invasive cells
form large invasive cones, protruding into the ECM. These invasive
cones are followed by weak protrusion of the proliferative rim,
leading to bumpy surface of the primary tumor. The fact that such
complex growth dynamics are only observed for tumors growing
heterogeneous ECM emphasizes the crucial importance of
understanding the effects of physical heterogeneity in cancer
research.

\section*{Discussion}




We have developed a novel cellular automaton (CA) model which, with just a few
parameters, can produce a rich spectrum of growth dynamics for invasive
tumors in heterogeneous host microenvironment. Besides robustly reproducing the salient features
of branched invasive growth, such as least resistance and intrabranch homotype attraction
observed in \textit{in vitro} experiments, our model also enables us
to systematically investigate the effects of microenvironment heterogeneity on
tumor growth as well as the coupling growth of the primary tumor and the invasive cells.
In particular, we have shown that in homogeneous ECM with low densities (i.e., soft microenvironment),
both the shape of the primary tumor and invasive pattern are isotropic. For high cellular
motility cases, the invasive cells form extended dendritic invasive branches; while for low
cellular motility cases, the invasive cells clump near the primary tumor surface
and form a bumpy concentric-like shell that facilitates the growth of the primary tumor.
Tumors growing in a highly rigid homogeneous ECM can develop anisotropic shapes,
facilitated by the invasive cells that degrade the ECM; both the tumor size and
the extent of invasive branches are much smaller.
In heterogeneous ECM, both the primary tumor and invasive pattern are
significantly affected during the early growth stages, i.e., anisotropic shapes
and patterns are developed to avoid high density/rigid regions of the ECM.
If the characteristic length scale of the heterogeneities is comparable to the macroscopic tumor size,
such effects can persist in later growth stages. In addition, invasive cells
with large motility can significant diminish the anisotropy effects by their
ECM degradation activities. We emphasize that we did not manipulate the behavior of cells
by imposing artificial CA rules to give rise to these complex and rich growth dynamics.
Instead, these are emergent behaviors that naturally arise due to
various microscopic-scale tumor-host interactions that are incorporated into our CA model,
including the short-range mechanical interaction between
tumor cells and tumor stroma, and the degradation of extracellular matrix by the
invasive cells.

Note that the growth dynamics of tumors growing in a heterogeneous
microenvironment is distinctly different than those in a
homogeneous microenvironment. This emphasizes the importance of
understanding the effects of physical heterogeneity of the host
microenvironment in modeling tumor growth. Here we just make a
first attempt to take into account a simple level of host
heterogeneity, i.e., by considering the ECM with variable
density/rigidity. Currently, the invasion of the malignant cells
into the host microenivronment is considered to be a consequence
of invasive cell phenotype gained by mutation, and is not
triggered by environmental stresses. However, the effects of
environmental stresses can be easily taken into account. For
example, a CA rule can be imposed that if the division probability
of a malignant cell is significantly reduced by ECM rigidity,
i.e., it is extremely difficult to push away/degrade ECM to make
room for daughter cells, the malignant cell leaves the primary
tumor and invades into soft regions of surrounding ECM. This would
lead to reduced tumor invasion (i.e., development of the dendritic
invasive branches) in soft microenvironments but enhanced invasion
in rigid microenvironments \cite{guiot07}.

Moreover, the spatial-temporal evolution of more complicated and
realistic nutrient/ oxygen fields can be incorporated into our CA
model. This can be achieved by solving the coupled nonlinear
partial differential equations governing the evolution of the
nutrient/oxygen concentrations as was done in
Refs.~\cite{gevertz06} and \cite{jana09}. Since the CA rules are
given for any spatial dimension, our model is readily generalized
to three dimensions. In addition, the model can be easily modified
to incorporate other host heterogeneities, such as stromal cells,
blood vessels and the shape anisotropy of the host organ
\cite{jana08, jana09}. As currently implemented, a single 2D
simulation takes less than 0.5 hours on a 32-bit 1.56Gb Memory
1.44GHz dual core Dell Workstation. We expect that a 3D simulation
will take no longer than 24 hours on a supercomputer when a proper
parallel implementation is used.

Such an \textit{in silica} tool not only enables one to investigate
tumor growth in complex heterogeneous microenvironment that closely represents
the real host microenvironments but also allows one to infer and even
reconstruct individual host microenvironment given limited growth data
of tumors (such as shape and size at various times). Such
microstructural information of the individual host would be extremely
valuable for developing individualized treatment strategies.
For example, based on the host microstructure one can design
special encapsulation and transport agents that maximize drug delivery efficiency \cite{To11}.

In our current CA model, the microscopic parameters governing
tumor invasion are variable and can be arbitrarily chosen within
a feasible range as given in Table \ref{tab_Param}. Given sufficient and reliable experimental data
of invasive tumor growth, the parameters in our CA model
can be uniquely determined and thus, the model can produce robust 
predictions on the neoplastic progression. 
Although the current CA model is specifically implemented to reproduce and 
predict the growth dynamics of invasive solid tumors \textit{in vitro},  
further refinement of the model could eventually lead to the development of a powerful
simulation tool that can be utilized clinically. For example, more complicated and
realistic host heterogeneities such as the vascular structure, various 
stromal cells, the corresponding spatial-temporal evolution of the 
nutrient/oxygen concentrations as well as environmental stress-induced 
mutations should be incorporated as we described earlier. 
If the robustness of the refined model could be validated clinically, we would 
expect it to produce quantitative predictions for {\it in vivo} tumor growth, which are
valuable for tumor prognosis and proposing individualized treatment strategies.




\section*{Acknowledgments}
The authors are grateful to Robert Gatenby and Bob Austin for
valuable comments on our manuscript. The authors are also 
grateful to the anonymous reviewers for their valuable comments.


\section*{Figure Legends}

\begin{table}[!ht]
\caption{\bf{Parameters and terms in the CA model}}
\begin{center}
\begin{tabular}{ll} \\ 
\multicolumn{2}{c}{\textbf{Time dependent terms}} \\ \hline
$L_t$ & Local tumor radius (varies with cell positions) \\ \hline
$L_{max}$ & Local maximum tumor extent (varies with cell positions) \\ \hline
$\delta_p$ & Characteristic proliferative rim thickness \\ \hline
$\delta_n$ & Characteristic living-cell rim thickness (determines necrotic fraction) \\ \hline
$p_{div}$ & Probability of division (varies with cell positions) \\ \hline
\\
\multicolumn{2}{c}{\textbf{Growth parameters}} \\ \hline
$p_0$ & Base probability of division, linked to cell-doubling time (0.192 and 0.384) \\ \hline
$a$ & Base necrotic thickness, controlled by nutritional needs ($0.58$ mm$^{1/2}$) \\ \hline
$b$ & Base proliferative thickness, controlled by nutritional needs ($0.30$ mm$^{1/2}$) \\ \hline
\\
\multicolumn{2}{c}{\textbf{Invasiveness parameters}} \\ \hline
$\gamma$ & Mutation rate (determines the number of invasive cells, 0.05) \\ \hline
$\chi$ & ECM degradation ability ($0.4-1.0$) \\ \hline
$\mu$ & Cell motility (the number of ``jumps'' from one automaton cell to another, $1-3$) \\ \hline
\\
\multicolumn{2}{c}{\textbf{Other terms}} \\ \hline
$\rho_{\mbox{\tiny{ECM}}}$ & ECM density (determines the ECM rigidity and varies with positions, $0.0-1.0$) \\ \hline
\\
\end{tabular}
\end{center}
\begin{flushleft} Summarized here are definitions of the parameters for tumor growth and invasion,
and all other (time-dependent) quantities used in the simulations.  For each parameter,
the number(s) listed in parentheses indicates the value or range of values
assigned to the parameters during the simulations. The values of the
parameters are chosen such that the CA model can reproduce reported
growth dynamics of GBM from the medical literature \cite{deisboeck01, kansal00a, jana08}.
\end{flushleft}
\label{tab_Param}
\end{table}


\begin{table}[!ht]
\caption{\bf{Comparison of tumor morphology metrics associated with
simulated MTS and experimental data at 24 hours after tumor initialization.}}
\begin{center}
\begin{tabular}{c@{\hspace{0.35cm}}c@{\hspace{0.35cm}}c}
\hline
Metrics & Simulated MTS & Experimental data  \\ \hline
Specific surface $s$ & 9.24 & 9.78  \\
Asphericity $\alpha$ & 1.09 & 1.12  \\
Angular anisotropy metric $\psi$ & 0.17 & 0.19 \\
\hline \\
\end{tabular}
\end{center}
\label{tab_MTS}
\end{table}

\begin{table}[!ht]
\caption{\bf{Morphology metrics for simulated tumors growing in homogeneous ECM.}}
\begin{center}
{\bf Noninvasive tumor in ECM with $\rho_{ECM} = 0.45$}\\
\begin{tabular}{c@{\hspace{1.75cm}}c@{\hspace{0.35cm}}c@{\hspace{0.35cm}}c@{\hspace{0.35cm}}c}
\hline
Metrics & Day 50 &  Day 80 &  Day 100 & Day 120  \\ \hline
Specific surface $s$ & 1.23 & 1.13 & 1.09 & 1.04 \\
Asphericity $\alpha$ & 1.21 & 1.18 & 1.08 & 1.12 \\
\hline \\
\end{tabular}

{\bf Invasive tumor with $\mu = 1$ in ECM with $\rho_{ECM} = 0.45$}\\
\begin{tabular}{c@{\hspace{0.35cm}}c@{\hspace{0.35cm}}c@{\hspace{0.35cm}}c@{\hspace{0.35cm}}c}
\hline
Metrics  & Day 50 &  Day 80 &  Day 100 & Day 120  \\ \hline
$\beta = A_I/A_T$ & 0.66 & 0.38 & 0.21 & 0.19 \\
Specific surface $s$ & 1.76 & 1.48 & 1.26 & 1.18 \\
Asphericity $\alpha$ & 1.23 & 1.12 & 1.08 & 1.06 \\
Angular anisotropy metric $\psi$ & 0.13 & 0.29 & 0.33 & 0.17 \\
\hline \\
\end{tabular}

{\bf Invasive tumor with $\mu = 2$ in ECM with $\rho_{ECM} = 0.45$}\\
\begin{tabular}{c@{\hspace{0.35cm}}c@{\hspace{0.35cm}}c@{\hspace{0.35cm}}c@{\hspace{0.35cm}}c}
\hline
Metrics & Day 50 &  Day 80 &  Day 100 & Day 120  \\ \hline
$\beta = A_I/A_T$ & 1.28 & 2.12 & 2.67 & 2.08 \\
Specific surface $s$ & 1.94 & 3.92 & 3.67 & 3.28 \\
Asphericity $\alpha$ & 1.42 & 1.38 & 1.16 & 1.23 \\
Angular anisotropy metric $\psi$ & 0.86 & 0.67 & 0.64 & 0.45 \\
\hline \\
\end{tabular}

{\bf Invasive tumor with $\mu = 3$ in ECM with $\rho_{ECM} = 0.45$}\\
\begin{tabular}{c@{\hspace{0.35cm}}c@{\hspace{0.35cm}}c@{\hspace{0.35cm}}c@{\hspace{0.35cm}}c}
\hline
Metrics & Day 50 &  Day 80 &  Day 100 & Day 120 \\ \hline
$\beta = A_I/A_T$ & 2.14 & 2.43 & 2.64 & 2.89 \\
Specific surface $s$ & 1.71 & 4.28 & 7.89 & 9.73 \\
Asphericity $\alpha$ & 1.38 & 1.27 & 1.13 & 1.8 \\
Angular anisotropy metric $\psi$ & 1.25 & 0.68 & 0.41 & 0.18 \\
\hline \\
\end{tabular}

{\bf Invasive tumor with $\mu = 3$ in ECM with $\rho_{ECM} = 0.85$}\\
\begin{tabular}{c@{\hspace{0.35cm}}c@{\hspace{0.35cm}}c@{\hspace{0.35cm}}c@{\hspace{0.35cm}}c}
\hline
Metrics & Day 50 &  Day 80 &  Day 100 & Day 120  \\ \hline
$\beta = A_I/A_T$ & - & 5.17 & 3.89 & 2.63 \\
Specific surface $s$ & 1.21 & 3.40 & 3.91 & 5.79 \\
Asphericity $\alpha$ & 1.33 & 1.36 & 1.40 & 1.56 \\
Angular anisotropy metric $\psi$ & - & 1.32 & 1.02 & 0.65 \\
\hline
\\
\end{tabular}
\end{center}
\label{tab_homo}
\end{table}

\begin{table}[!ht]
\caption{\bf{Morphology metrics for simulated tumors growing in heterogeneous ECM.}}
\begin{center}
{\bf Invasive tumor with $\mu = 3$ in random ECM}\\
\begin{tabular}{c@{\hspace{0.35cm}}c@{\hspace{0.35cm}}c@{\hspace{0.35cm}}c@{\hspace{0.35cm}}c}
\hline
Metrics & Day 50 &  Day 80 &  Day 100 & Day 120  \\ \hline
$\beta = A_I/A_T$ & 2.54 & 4.14 & 2.13 & 2.78 \\
Specific surface $s$ & 2.47 & 3.97 & 4.65 & 8.98 \\
Asphericity $\alpha$ & 1.32 & 1.34 & 1.18 & 1.15 \\
Angular anisotropy metric $\psi$ & 0.63 & 0.87 & 0.64 & 0.19 \\
\hline \\
\end{tabular}

{\bf Invasive tumor with $\mu = 1$ in sine-like ECM}\\
\begin{tabular}{c@{\hspace{0.35cm}}c@{\hspace{0.35cm}}c@{\hspace{0.35cm}}c@{\hspace{0.35cm}}c}
\hline
Metrics & Day 50 &  Day 80 &  Day 100 & Day 120  \\ \hline
$\beta = A_I/A_T$ & 2.78 & 2.46 & 1.28 & 0.86 \\
Specific surface $s$ & 1.89 & 2.95 & 2.73 & 1.92 \\
Asphericity $\alpha$ & 1.42 & 1.61 & 1.49 & 1.26 \\
Angular anisotropy metric $\psi$ & 1.23 & 1.18 & 1.09 & 0.98 \\
\hline \\
\end{tabular}

{\bf Invasive tumor with $\mu = 3$ in sine-like ECM}\\
\begin{tabular}{c@{\hspace{0.35cm}}c@{\hspace{0.35cm}}c@{\hspace{0.35cm}}c@{\hspace{0.35cm}}c}
\hline
Metrics & Day 50 &  Day 80 &  Day 100 & Day 120   \\ \hline
$\beta = A_I/A_T$ & 5.24 & 3.86 & 3.13 & 2.96 \\
Specific surface $s$ & 2.51 & 4.12 & 6.13 & 8.76 \\
Asphericity $\alpha$ & 1.19 & 1.67 & 1.48 & 1.21 \\
Angular anisotropy metric $\psi$ & 1.36 & 1.33 & 0.88 & 0.23 \\
\hline
\\
\end{tabular}
\end{center}
\label{tab_hetero}
\end{table}

\end{document}